\def\build#1_#2^#3{\mathrel{
\mathop{\kern0pt#1}\limits_{#2}^{#3}}}

\def\bar{\overline}
\def\ben{\begin{equation}}
\def\be#1{\begin{equation}\label{eq:#1}}
\def\ee{\end{equation}}
\def\EC#1{(\ref{eq:#1})}
\def\reff{\par\noindent\hangafter=1\hangindent=1cm}
\def\bibitem{\reff}
\documentstyle[11pt,aaspp4,fleqn,epsf]{article}
\def\build#1_#2^#3{\mathrel{
\mathop{\kern0pt#1}\limits_{#2}^{#3}}}
\def\la{\mathrel{\mathpalette\fun <}}

\def\fun#1#2{\lower3.6pt\vbox{\baselineskip0pt\lineskip.9pt
        \ialign{$\mathsurround=0pt#1\hfill##\hfil$\crcr#2\crcr\sim\crcr}}}

\def\bar{\overline}

\def\ben{\begin{equation}}
\def\be#1{\begin{equation}\label{eq:#1}}
\def\bea#1{\begin{eqnarray}\label{eq:#1}}
\def\ee{\end{equation}}
\def\eea{\end{eqnarray}}
\def\EC#1{(\ref{eq:#1})}

\def\reff{\par\noindent\hangafter=1\hangindent=1cm}          

\begin{document}

\begin{titlepage}
\null\vspace{-62pt}
\vspace{1.5in}
\baselineskip 12pt
\centerline{{\Large \bf Relaxing and Virializing a Dark Matter Halo}}
\vspace{0.5in}
\centerline{{\bf Richard N. Henriksen 
\footnote{E-mail address: henriksn@astro.queensu.ca}}
and {\bf Lawrence M. Widrow
\footnote{E-mail address: widrow@astro.queensu.ca}}}
\vspace{0.1in}
\centerline{{\it Department of Physics}}
\centerline{{\it Queen's University, Kingston, K7L 3N6, CANADA}}

\begin{abstract}

Navarro, Frenk, and White have suggested that the density profiles of
simulated dark matter halos have a ``universal'' shape so that a given
halo can be characterized by a single free parameter which fixes its
mass.  In this paper, we revisit the spherical infall model in the
hope of recognizing in detail the existence and origin of any such
universality.  A system of particles is followed from linear
perturbation, through first shell crossing, then through an accretion
or infall phase, and finally to virialization.  During the accretion
phase, the system relaxes through a combination of phase mixing, phase
space instability, and moderate violent relation.  It is driven
quickly, by the flow of mass through its surface, toward self-similar
evolution.  The self-similar solution plays its usual r\^ole of
intermediate attractor and can be recognized from a virial-type
theorem in scaled variables and from our numerical simulations.  The
transition to final equilibrium state once infall has ceased is
relatively gentle, an observation which leads to an approximate form
for the distribution function of the final system.  The infall phase
fixes the density profile in intermediate regions of the halo to be
close to $r^{-2}$.  We make contact with the standard hierarchical
clustering scenario and explain how modifications of the self-similar
infall model might lead to density profiles in agreement with those
found in numerical simulations.

\end{abstract}

\end{titlepage}

\newpage

\section{Introduction}

Navarro, Frenk and White (1996) have summarized the results of their
dissipationless cosmological clustering simulations in terms of a
`universal' shape for the density profile of dark halos.  This profile
(henceforth the NFW profile),

\be{nfw_profile}
\rho(r)~=~\frac{M_s}{r\left (r+R_s\right )^2}~,
\ee

\noindent is characterized by an $r^{-1}$ central cusp and an outer 
region where the density falls off faster than that of an isothermal
sphere.  Good fits are obtained using Eq.\,\EC{nfw_profile} for halos
that range in mass from $3\times 10^{11} M_\odot$ (dwarf galaxies) to
$3\times 10^{15}M_\odot$ (rich galaxy clusters).  In addition, the
results show a strong correlation between the constants $R_s$ and
$M_s$ so that there is essentially a single free parameter (Navarro,
Frenk, \& White 1996).  The possible existence of a universal profile
suggests that the structure of collapsed objects can be understood
from simple physical arguments.

The earliest attempts to understand cosmological structure were based
on the spherical radial infall model (e.g., Gunn \& Gott 1972;
Henriksen \& De Robertis 1980; Fillmore \& Goldreich 1984, hereafter
FG; Bertschinger 1985, hereafter B85; Hoffman \& Shaham 1985; White \&
Zaritsky 1992) in which a primordial density perturbation, assumed to
be spherically symmetric and smooth, slowly accretes matter from the
cosmic background.  If this initial perturbation is also scale-free
(i.e., $\delta\rho_i\propto r^{-\epsilon}$, see e.g., FG) with a
velocity distribution corresponding to an unperturbed Hubble law, the
structure that develops will be self-similar in the sense that the
distribution function at one time can be obtained from that at another
time by rescaling the phase space coordinates $r$ and $v_r$. One way
to determine the scaling law is to note that at each time $t$ there is
a single mass shell that is just beginning to break away from the
expansion and fall in towards the center.  The radius at which this
occurs defines a function of time, $r_{\rm ta}\propto t^{\delta}$
where

\be{simclass} 
\delta= \frac{2}{3} + \frac{2}{3\epsilon}~,
\ee 

\noindent suggesting that the appropriate radial coordinate for 
the self-similar solution is $X\equiv r/r_{\rm ta}$ (FG and B85).

Each FG and B85 solution represents a single trajectory in phase
space, albeit one that describes multiple velocity streams in the
inner parts of the system.  An alternative approach (Henriksen \&
Widrow 1997, hereafter HW), based directly on the collisionless
Boltzmann equation (CBE), treats the distribution function as
continuous in the scaled phase-space variables.  In this picture, the
single-trajectory solutions of FG and B85 represent a subset of the
characteristic curves of the CBE.  Numerical simulations of systems
that begin from a ``cold start'' (i.e., single-stream distribution
function) show a rapid transition to a distribution function that is
continuous, the transition being facilitated by an instability (HW).
The form of the scaling in the CBE (dictated by $\delta$) is
equivalent to the similarity `class' in the sense of Carter and
Henriksen (1991).  This is determined by the logarithmic derivative of
$r_{\rm ta}$ with respect to future turn-round time {\it at the epoch
of first shell crossing}, and is given by Eq.\EC{simclass}.

Self-similarity would seem to be an appealing feature for a model that
is to explain universal characteristics in cosmological structures of
the type described by Navarro, Frenk, \& White (1996).  This is
particularly so given the usual r\^ole of self-similarity as an
attractor.  However, the self-similar infall model (SSIM) has been, by
and large, dismissed as a paradigm for structure formation.  First,
the SSIM does not include angular momentum which likely plays a key
role in determining the density profile, at least in the central
regions.  Second, structure formation is thought to develop by way of
hierarchical clustering: Small mass objects form first and merge with
one another to create systems of ever-increasing size.  This process
is neither smooth nor spherically symmetric.  Moreover, the solutions
found by FG and B85 describe an eternal infall of matter and say
nothing about how a system might enter or exit such a self-similar
state.  In the standard scenario for structure formation,
gravitational collapse begins only after the Universe has become
matter dominated.  Moreover, at late times (e.g., as for galaxies
today), the evolution of a system is dominated by infrequent mergers
with comparable-sized objects.  Therefore at best, self-similarity
will arise as an intermediate phase in the evolution of a system.
(This is of course always the case with self-similar behaviour.)
Finally, it is generally believed that the SSIM predicts a power-law
density profile for ``virialized'' halos, ($\rho(r)\sim r^{-\mu}$ with
$2<\mu<9/4$ where $\mu$ depends on $\epsilon$) in contrast with what
is found in the simulations (cf. Eq.\,\EC{nfw_profile}).

In this paper we argue instead that the SSIM provides a natural
framework for understanding cosmological structure formation, A simple
shell code is used to treat spherically symmetric collisionless
particles (i.e., spherical shells) on radial orbits.  The particles
are followed from linear perturbation through first shell crossing,
then through the recently recognized self-similar relaxation phase,
and into eventual virialization which occurs after the cessation of
infall.  Many features of the self-similar solutions discovered by FG
and B85 are present in this intermediate phase though much of the
fine-grained phase-space structure characteristic of these solutions
is washed out.  We maintain that there is a close connection between
the SSIM and semi-analytic models of structure formation based more
directly on hierarchical clustering (e.g., Lacey \& Cole 1993).  In
addition we will show that simple modifications of the SSIM lead to
density profiles in agreement with results from N-body simulations.
An advantage of this approach is that it provides a connection,
through a simple semi-analytic model, between initial conditions and
the distribution function and density profile of the final virialized
system.

In a hierarchical clustering universe, the progenitors of present day
nonlinear structures are small-amplitude primordial density
fluctuations.  Bond et al.\,(1991) and Lacey \& Cole (1993) have
developed an analytic model for hierarchical structure formation that
relates the initial power spectrum to the mass and formation time of
dark matter halos.  In their formalism, the linear density field at
some early time, $t_i$, is smoothed on various mass scales.  Objects
that will collapse by some later time, $t_{\rm coll}$, are identified
at the earlier epoch as regions where the mean density is above a
certain threshold.  This threshold is estimated using the spherical
top-hat model, i.e., the perturbation within a given radius is modeled
as a region of constant density.  Of course in the spherical top-hat
model, all parts of the perturbation collapse to the center at the
same time and it is therefore necessary to make an {\it ad hoc}
assumption about what happens after collapse.  Lacey \& Cole (1993)
assume that at $t_{\rm coll}$ the system reaches virial equilibrium
with a radius equal to half its maximum or turnaround radius.  Given
this assumption, the mean density within a collapsed object at a
particular epoch turns out to be roughly $\sim 200$ times the
background density at that time.  It is now common practice to define
the `virial radius', $r_{v}$, as the radius of a sphere that contains
an overdense region whose mean density is $v$ times the background
density.  The claim is that the mass within this sphere has, more or
less, reached a final equilibrium state.  $r_v$ is used, for example,
to identify virialized halos in N-body simulations (e.g., Cole \&
Lacey 1996, Navarro, Frenk, \& White 1996).

Let us see how the SSIM might improve this picture.
If the initial perturbation spectrum is
scale-free ($\langle |\delta_k |^2\rangle\propto k^n$) the RMS mass
fluctuation will be a power-law function of radius, $\langle
\delta_R({\bf x})^2\rangle^{1/2}\propto R^{-(n+3)/2}$.  In this
expression, $R$ is the radius of the window function used in computing
the mass fluctuation, ${\bf x}$ is the position vector about which
this window function is centered, and $\langle\dots\rangle$ indicates
an average over ${\bf x}$.  This suggests an initial density
perturbation with $\delta\rho_i\propto r^{-\epsilon}$
($\epsilon=(n+3)/2$) as the most appropriate ``toy-model'' for
understanding structure formation via hierarchical clustering and allows
us to write the similarity class $\delta$ and final density profile
index $\mu$ in terms of the spectral 
index $n$:

\be{deltavsn}
\delta = \frac{2}{3}\left (\frac{n+5}{n+3}\right )~;
\ee

\be{muvsn}
\mu=\left\{ 
        \begin{array}{ll}
3\left (\frac{n+3}{n+5}\right )&\mbox{$n>1$} \\
\\
2&\mbox{$n\le 1$} 
\ \ \ .
        \end{array} 
        \right.
\ee 

\noindent Indeed the results obtained from the SSIM are 
consistent with those found in Lacey \& Cole (1993). Their {\it ad
hoc} virialization radius $r_{v}$ corresponds to an $X={\rm constant}$
$\left (r={\rm constant}\times t^\delta\right )$ surface and the
expression for the mass within this surface, as a function of $t$,
leads to their result for the mass-formation time relationship.  While
the ``scaled mass'' within $r_{v}$ approaches a constant, the physical
mass increases as $t^{2/\epsilon}=t^{3\delta-2}$.  A mass flux through
the system boundary is required but this is built into the
self-similar solution.  The SSIM avoids the singular nature of the
collapse in the spherical top-hat model and thereby affords insight
into the virialization process and the ultimate fate of the halo once
the mass flux has ceased.

But the question remains as to the relevance of the SSIM to halo
formation via hierarchical clustering.  Recently, Syer and White
(1997) (see, also Nusser and Sheth, 1997) have suggested that the
universality seen in the simulations reflects a balance between two
opposing processes, dynamical friction and tidal stripping, that act
on a small object as it merges with a larger one.  Dynamical friction
will bring a satellite to the centre of a parent halo but only if the
satellite has not been tidally disrupted.  Whether or not the
satellite reaches the centre intact depends on the density profile of
the parent.  For a steep density profile, the satellite is disrupted
before it reaches the centre so that its material is spread throughout
the halo, thus softening the density law.  Conversely, if the density
profile of the parent halo is relatively flat, the satellite reaches
the centre largely intact and so boosts the density there.  In this
way, dynamical friction and tidal dissipation act as negative feedback
mechanisms driving the density profile toward a universal shape (Syer
\& White 1997).  (For an alternative discussion in which the 
universality of dark halo density profiles does not depend crucially
on hierarchical merging, see Huss, Jain, \& Steinmetz 1998.)

A similar situation arises in the SSIM (FG, Moutarde et al.\,1995).
Suppose the initial density perturbation is relatively flat (here flat
and steep are with respect to a $\rho\propto r^{-2}$ density law).
The system evolves toward a universal profile which, in the
intermediate regions of the halo, is $\propto r^{-2}$ and is a
self-similar attractor which we discuss in detail below. As in the
hierarchical scenario, the central regions are dominated by material
that has fallen in recently (FG) since the binding energy, $GM(r)/r$,
is an increasing function of $r$.  By contrast, steep initial density
profiles evolve stably as one of a 1-parameter continuum of
self-similar solutions.  The parameter is the quantity $\delta$
introduced above (related ultimately to $n$) and the density profile
is the same as that of the perturbation at first shell crossing.  The
$\rho\propto r^{-2}$ attractor is actually the flattest of these
self-similar profiles.  We can say then that the universality
predicted by the SSIM is `one-sided'.  It is worth mentioning however
that the density laws for the stably evolving systems in the SSIM vary
only from $r^{-2}$ to $r^{-9/4}$ for $n\in\{1,3\}$.

As discussed above, a major criticism of the SSIM is that the density
profile it predicts for collapsed objects does not agree with the
results from N-body simulations.  The power-law index in the NFW
profile, for example, varies smoothly from $-1$ in the core to $-3$ in
the envelope, in contrast with what is found in the similarity
solutions discussed above.  There is some controversy over what the
true density law within the inner regions of simulated halos is.
Moore et al.\,(1997) find that mass resolution and force softening
have a significant effect on the density profiles of collapsed objects
in collisionless N-body simulations.  As the mass and force resolution
is increased the density profile in the central region becomes
steeper.  Even with 3 million particles per halo, the results have not
converged to a unique density profile.  Moore et al.\,(1997) attribute
their results to the issue of dynamic range in the clustering
hierarchy: With better resolution, smaller halos would collapse
earlier causing the density profile to steepen.  A similar situation
arises in the SSIM.  As noted above, the solutions derived by FG and
B85 correspond to an eternal infall of matter on a single trajectory.
If instead, collapse begins at a finite time, there would be
relatively fewer particles in the inner regions leading to a shallower
density law in the center.  In the real Universe, the size of the
first objects to form is set by the horizon size at matter-radiation
equality while in the simulations it is set by the force and mass
resolution of the experiment.

As noted above, the NFW profile is noticeably steeper than $r^{-2}$ in
the outer regions of simulated halos.  We propose that this region
forms after the primary accretion phase when the evolution of a system
becomes dominated by major mergers.  The system is still accreting
mass during this phase, but not at a rate sufficient to maintain
self-similar growth.  The infalling material can be treated as test
particles and naturally forms an $r^{-3}$ outer halo.

In Section 2 we write the basic equations describing collisionless
radial dynamics (CBE, mass conservation and Euler equations,
virialization condition) in scaled variables.  We use these variables
to follow numerically a collisionless spherical cosmological
perturbation from linear perturbation to a self-similar infall phase
(steady-state in scaled variables) and finally to a true virialized
state which arises once the stream of particles falling into the
perturbation is shut off.  The simulations are presented in Section 3.
Section 4 presents a more detailed discussion of the transition from
infall phase to virialized isolated system.  A summary and conclusions
are given in Section 5.

\section{`Virialization' in Scaled Variables}

We begin this section by writing the CB and Poisson equations in
scaled variables.  Standard manipulations (see, for example Binney \&
Tremaine 1987) lead to the usual mass conservation and Euler equations
as well as the virial theorem in these variables.  This formulation is
inspired by the analysis of self-similarity given by Carter \& Henriksen
(1991) (see also Henriksen (1997) for a more physical
exposition and Henriksen \& Widrow (1995, 1997) for relevant
examples).  A self-similar state is identified as one that is
independent of $T$.  $X$ and $Y$ will then correspond to the
self-similar or scale-invariant variables.  Therein lies the interest
in this formulation, since the emergence of a time-independent state
in these variables allows one to recognize self-similarity of the type
found in FG, B85, Moutarde et al.\,1995, HW, and others.

The CB and Poisson equations for a spherically symmetric system are

\be{sphericalv}
\frac{\partial f}{\partial t} + v_r
\frac{\partial f}{\partial r} +\left({j^2\over r^3}-
\frac{\partial\Phi}{\partial r}\right)
\frac{\partial f}{\partial v_r}~=~0~,
\ee

\be{sphericalp}
\frac{\partial}{\partial r}
\left (r^2\frac{\partial\Phi}{\partial r}\right )~=~
4\pi^2\int f(r,v_r,j^2)\,dv_r dj^2
\ee

\noindent where $f$ is the phase space mass density and $\Phi$ is the
`mean field' gravitational potential.  (We are considering scales much
smaller than the Hubble length.)  For purely radial orbits it is
convenient to introduce the canonical distribution function
$F(t,r,v_r)$ where $f\equiv \left (4\pi^2\right )^{-1}\delta(j^2)F$.
The equations then become

\be{radialv}
\frac{\partial F}{\partial t} + v_r
\frac{\partial F}{\partial r} -
\frac{\partial\Phi}{\partial r}
\frac{\partial F}{\partial v_r}~=~0~,
\ee

\be{radialp}
\frac{\partial}{\partial r}\left ( 
r^2\frac{\partial\Phi}{\partial r}\right )~=~G\int F dv_r~.
\ee
 
\noindent The next step is to introduce the scaled phase-space 
variables

\be{scaled_r}
X \equiv e^{-\delta T}r
\ee

\be{scaled_v}
Y \equiv e^{(1-\delta) T}v_r
\ee

\noindent and

\be{scaled_t}
T \equiv \ln(t/t_i)~,
\ee

\noindent along with the appropriately scaled distribution
function and potential:

\be{scaled_df}
{\cal F}(X,Y,T)=Ge^{(1-\delta)T}F(r,v_r,t)
\ee

\be{scaled_phi}
{\Psi}(X,Y,T)=e^{2(1-\delta) T}\Phi(r,v_r,t)~.
\ee

\noindent The CB and Poisson equations become

\be{scale_cbe}
\frac{\partial{\cal F}}{\partial T} + 
\left (\delta -1\right ){\cal F} + 
\left (Y-\delta X\right )\frac{\partial{\cal F}}{\partial X}
- \left (\left (\delta -1\right )Y + \frac{\partial{\Psi}}
{\partial X}\right )
\frac{\partial {\cal F}}{\partial Y} = 0
\ee

\be{scale_poisson}
\frac{\partial}{\partial X}
\left (X^2\frac{\partial{\Psi}}{\partial X}\right )=\int {\cal F} dY~.
\ee

\noindent Integrating Eq.\,\EC{scale_cbe} 
over $Y$ yields the continuity equation:

\be{continuity}
\frac{\partial{\cal S}}{\partial T}+
\left (3\delta - 2\right){\cal S}+
\frac{\partial}{\partial X}
\left ({\cal S}\left ({\bar Y}-\delta X\right )\right )=0
\ee

\noindent where 

\be{def_moments}
{\cal S}\equiv\int dY{\cal F}~~~~~~;~~~~~~~
{\cal S}{\bar Y}\equiv\int dY Y{\cal F}~.
\ee

\noindent ${\cal S}$ is essentially the scaled mass per unit radius
and is related to the physical density: $\rho\equiv\int d^3v f =
\exp{(-2T)}{\cal S}/4\pi GX^2$.

The mass conservation equation is obtained by integrating
Eq.\,\EC{continuity} over $X$. (Note that we have assumed ${\cal
S}\rightarrow 0$ at $X\rightarrow 0$.)  We find

\be{mass_cons}
\frac{d{\cal M}}{dT}+\left (3\delta -2\right ){\cal M}
+\left\{{\cal S}\left ({\bar Y}-\delta X\right )\right\}_{X=X_{\rm s}}=0
\ee

\noindent where $X_{\rm s}$ defines the (spherical) boundary of the system and

\be{def_mass}
{\cal M}\equiv \int_0^{X_{\rm s}}{\cal S} dX = \int_0^{X_{\rm s}}
dX\int_{-\infty}^\infty dY{\cal F}
\ee

\noindent is the scaled mass within this boundary, related to the 
physical mass $M$ by

\be{phys_mass}
M=e^{\left (3\delta-2\right )T}{\cal M}
=\left (\frac{t}{t_i}\right )^{\left (3\delta-2\right )}{\cal M}~.
\ee

\noindent For an isolated system, ${\cal S}\to 0$ for $X\to\infty$.  Taking
$X_{\rm s}\to\infty$ we find (from \EC{mass_cons}) ${\cal M}\propto
e^{-\left (3\delta-2\right )T}$ which states the obvious: the mass of
an isolated system is constant.

In a cosmological setting, the distribution of particles extends to 
$X=\infty$.  A particularly simple example of a self-similar model
is the matter-dominated Einstein-de Sitter universe which is described
in terms of our scaled variables as follows:

\be{EdSdf}
{\cal F} = \frac{2\rho_i}{3}X^2\delta_D\left (Y-\frac{2X}{3}\right )
\ee

\be{EdSdensity}
{\cal S} = \frac{2\rho_i}{3}X^2~~~~~~~~~~~~~~
\left ({\rm equivalently},~~~\rho = \frac{1}{6\pi Gt^2}\right )
\ee

\be{EdSybar}
{\bar Y} = \frac{2X}{3}  ~~~~~~~~~~~~~~\left ({\rm equivalently},~~~
v_r = \frac{2r}{3t}\right )
\ee

\noindent where $\delta_D$ is the Dirac delta function.
The similarity class is characterized by $\delta = 2/3$
with $X$ being the usual comoving radius.  However, by adding
a power-law perturbation to this background density, one changes the
similarity class and obtains the models considered in FG and B85 as
well as in the present work.

A self-similar state requires by definition that the quantities ${\cal
S}$, ${\cal M}$, and ${\bar Y}$ be independent of $T$.  We conclude
from Eq.\,\EC{mass_cons} that the surface term $\left\{{\cal S}\left
({\bar Y}-\delta X\right )\right\}_{X=X_{\rm s}}$ must also be independent
of $T$ {\it and} nonzero, unless $\delta=2/3$. The continuing infall
thus {\it maintains} the self-similarity.

Ultimately in these calculations the similarity class, $\delta$, is
set by the density profile of the initial perturbation.  Thus for
$\delta\rho_i\propto r^{-\epsilon}$, at the epoch of first shell
crossing the initial shell label or `effective radius' of a shell (see
e.g. Henriksen 1989) that will turn around at time $t$ is
asymptotically proportional to $t^{2/3+2/3\epsilon}$. 
This establishes a fundamental relation between the time and
space scaling in the subsequent infall (we have set the time scaling
equal to unity) which suggests that we set
$\delta=2/3+2/3\epsilon$. The requirement for the system boundary to
be fixed at $X_{\rm s}$ in the self-similar state, as found above, implies
that the current boundary in physical coordinates varies as
$t^{\delta}$. This boundary is approximately the current turn-round
radius $r_{\rm ta}$ which then varies also as $t^{\delta}$ in
agreement with the arguments of FG and B85.

In the general self-similar state we have, from Eq.\,\EC{mass_cons},

\be{mass_ss}
{\cal M}=\left (3\delta -2\right )^{-1}\left\{{\cal S}
\left (\delta X - {\bar Y}\right )\right\}_{X=X_S}
\ee

\noindent in agreement with the semi-analytic results of B85
(cf. his Eq.(4.5)).  Eq.\,\EC{mass_ss}, together with
Eq.\,\EC{phys_mass}, imply that $M\propto t^{(3\delta-2)T}\propto
t^{2/\epsilon}$.  In other words, the physical mass of the system
increases with time due to a mass flux through the system boundary.
To make the connection with previous work on hierarchical clustering,
consider initial conditions in which the power spectrum is a pure
power law, $\langle |\delta_k |^2\rangle\propto k^n$.  The RMS mass
fluctuation as a function of scale (essentially the radius of a
spherical window function) is $\delta M/M\propto R^{-(n+3)/2}$.  Let
us suppose that our initial spherical perturbation has the same
dependence on radius as the RMS mass fluctuation derived from Gaussian
random fields, i.e., $\epsilon = (n+3)/2$.  With this identification,
we can relate the mass $M$ of a relaxing system to the time $t$, or
equivalently redshift $z$.  We find $1+z\propto M^{\left (n+3\right
)/6}$, in agreement with Lacey \& Cole (1993)

Multiplying Eq.\,\EC{scale_cbe} by $Y$ and integrating over
$Y$ leads to the Jeans equation in scaled variables:

\be{euler}
\frac{\partial\left ({\cal S}{\bar Y}\right )}{\partial T}+
3(\delta -1){\cal S}{\bar Y}-
\delta X\frac{\partial\left ({\cal S}{\bar Y}\right )}{\partial X}+
\frac{\partial\left ({\cal S}{\bar {Y^2}}\right )}{\partial X}+
{\cal S}\frac{\partial\Psi}{\partial X}=0~.
\ee

\noindent where 

\be{def_moments2}
{\cal S}{\bar {Y^2}}\equiv\int dY Y^2{\cal F}~.
\ee

\noindent A virial theorem in scaled variables is 
obtained by multiplying Eq.\EC{euler} by $X$ and integrating over $X$:

\be{virial}
\frac{1}{2}\frac{d^2{\cal I}}{dT^2} + \frac{5}{2}\left (2\delta-1\right )
\frac{d{\cal I}}{dT}+\frac{\left (5\delta-2\right )\left (5\delta-3\right )}
{2}{\cal I} + \left (5\delta-3\right ){\cal A}-\frac{d{\cal A}}{dT}
-2{\cal K}-{\cal W}=0
\ee

\noindent where 

\be{kinetic_potential}
{\cal K}\equiv\frac{1}{2}\int_0^{X_S}{\cal S}{\bar {Y^2}} dX~~~~~~~~
{\cal W}\equiv-\int_0^{X_S}{\cal S}X\frac{\partial\Psi}{\partial X} dX
~~~~~~~{\cal I}=\int X^2 {\cal S} dX
\ee

\noindent are the scaled total kinetic and potential energies and
moment of inertia of the system 
respectively and 

\be{surface}
{\cal A}\equiv \left\{{\cal S}
X\left ({\bar {Y^2}} -\delta X {\bar Y}\right ) \right \}_{X=X_S}
\ee

\noindent is a surface term.  In the self-similar state we have

\be{virial2}
\frac{2{\cal K}}{{\cal W}} - 1 = \frac{\left (5\delta-2\right )\left
(5\delta-3\right )} {2}\frac{{\cal I}}{{\cal W}} +
\left (5\delta-3\right )\frac{\cal A}{\cal W}~.
\ee

\noindent This virial theorem is satisfied for a system that is 
undergoing self-similar infall.  The left-hand side is reminiscent
of the usual virial theorem (e.g., Binney \& Tremaine, 1987).  The
first of the two terms on the right-hand side comes from the inherent
time-dependence of the infall solution and is analogous to the 
terms found in cosmology when working in comoving coordinates.  The
second term is due to the infall of matter through the system boundary.
We can of course rewrite this equation in terms of more 
familiar quantities such as the total kinetic energy,
$K=2\pi\int\rho {\bar {v_r^2}} r^2 dr = {\cal K}\exp{\left (5\delta -4
\right )T}$.  The result has essentially the same form, i.e., 
$2K/W={\it constant}$ and suggests that we consider the self-similar
infall phase as a type of time-dependent virial equilibrium.

\section{Numerical Simulations}

The FG and B85 solutions describe an eternal system in which
self-similarity is exact.  The profile of the initial density
perturbation is used to determine the appropriate scaling relations
but is otherwise not part of the solution, i.e., the self-similar
solutions {\it do not} follow a system from initial perturbation to
self-similar phase.  Nor, for that matter, do they shed light on the
ultimate state of the system once infall has ceased.  The simulations
described in this section are designed to address these issues.

A simple shell model is used to follow the evolution of spherically
symmetric density perturbations in an Einstein-de Sitter universe.
This model describes a system of collisionless, self-gravitating
``particles'' each of which represents a spherical shell of matter.
The scale of the perturbations is assumed to be small as compared with
the Hubble length.  The equations of motion for a shell of radius
$r(t)$ and radial velocity $v_r(t)$ are therefore given by Newtonian
dynamics:

\be{eofm_nbody}
\frac{dr}{dt} = v_r~~~~~~~~~~~~~~~
\frac{dv_r}{dt} = -\frac{GM(r,t)}{r^2}~.
\ee

\noindent At some initial time $t_i$, the unperturbed Hubble-flow is 
described by a constant background density, $\rho_b(t_i)=1/6\pi G
t_i^2$, and a velocity field $v_r(t_i)=2r(t_i)/3t_i$.  We introduce a
cut-off in the initial mass distribution at a finite radius $r_0$.
This cut-off will allow us to follow the evolution from infall phase
to isolated system.  The mass of a given shell is chosen to be
proportional to the square of its radius at time $t=t_i$: i.e.,
$m=2r^2(t_i)r_0/3NGt_i^2$ where $N$ is the number of shells.  In other
words, at $t=t_i$, the total mass in the system is divided into shells
of uniform thickness $r_0/N$.  By partitioning the mass in this way,
we improve the resolution of the simulation during the early stages of
collapse but at the cost of having a system which is modelled by
unequal masses.

The initial conditions that lead to the self-similar solutions of FG
and B85 assume a density profile that is a power-law function of
radius: $\delta\rho_i\propto r^{-\epsilon}$.  As discussed above, this
choice can be justified for models where the spectrum of density
perturbations is a scale-free function of wavenumber.  However there
are various reasons to expect that actual perturbations will be
modified at small scales.  First, even if the primordial perturbation
spectrum is scale-free, physical processes (e.g., neutrino
free-streaming, radiation damping) will suppress small-scale
fluctuations.  Second, the connection between the primordial power
spectrum and expected rms mass excess is valid only for linear
perturbations and breaks down near nonlinear density peaks.  There
are additional effects, peculiar to N-body simulations, related to
finite mass and spatial resolution, which also damp small scale
fluctuations.

These considerations suggest an initial density perturbation of the 
form

\be{rho_init}
\rho(r,t_i) = \rho_b(t_i)\left (
1+\Delta(r, t_i)\right )
\ee

\noindent where

\be{pert_def} 
\Delta(r,t_i)=\left\{ 
        \begin{array}{ll}
A\left (1-B\left (r/r_c\right )^2\right )&\mbox{$r<r_c$} \\
\\
A\left (1-B\right )\left (r/r_c\right )^{-\epsilon}&\mbox{$r\ge r_c$} 
\ \ \ .
        \end{array} 
        \right.
\ee 

\noindent We set $B=5\epsilon/\left (3\epsilon + 6\right )$ 
so that, for $r\ge r_c$, the mass interior to $r$ is what it would
have been had the density profile been a strict power-law function of
radius.  (This choice is made to maintain a close connection with the
FG solutions.  Other forms for the initial density profile are
possible.  See, for example Hoffman \& Shaham 1985; Bardeen, Bond,
Kaiser, \& Szalay 1986; Ryden \& Gunn, 1987; and Ryden, 1988.)
 
The desired initial density perturbation is achieved by
displacing each shell by an amount $\delta r(r_i)$ where

\be{perturb}
\frac{\delta r(r_i)}{r_i} = -\frac{1}{r_i^3}\int_0^{r_i} dr r^2
\Delta\left (r,t_i\right )
\ee

\noindent A modified force law 
($dv_r/dt = -GM(r,t)r/\left (r^2+\varepsilon^2\right )^{3/2}$) is used
to handle the dynamics of shells as they pass through the origin.

As discussed in Section II, the natural variables for a system of this
type are $X, Y$ and $T$.  We therefore perform our 
simulations in these variables.  Eqs.\,\EC{eofm_nbody} become:

\be{eofm_scaled}
\frac{dX}{dT}=Y-\delta X~~~~~~~~~~~~
\frac{dY}{dT} = \left (1-\delta\right )Y-
\frac{{\cal M}(X,T)}{X^2}
\ee
\noindent where ${\cal M}=\exp{\left (2-3\delta\right )}\times
\left ({\rm mass~interior~to~}X\right )$.  These are just the 
equations one would write to describe the characteristic curves of the
CBE (see HW and Eq.\,\EC{scale_cbe}).  At the initial time $t_i$,
$T=0$, $r=X$, and $v_r=Y$.

Gravitational collapse simulations are performed for a variety of
parameters.  In Figure 1 we show the evolution of a system of 16,000
particles in $(X,Y)$ phase space.  The density profile of the initial
perturbation is given by Eqs.\,\EC{rho_init} and \EC{pert_def} with
$\epsilon=2.5$.  This implies a similarity class characterized by
$\delta=14/15$.  The panels a-d in Figure 1 show the system at
$T=2.7,\, 8.0,\, 13.3$ and $18.7$ or equivalently $t/t_i=14,\,
3.0\times 10^3,\, 6.2\times 10^5,$ and $1.3\times 10^8$ respectively.
If the final frame is taken as the present epoch, then the initial
frame corresponds to a redshift of $z_i=2.6\times 10^5$.  As expected,
the size of the system in scaled variables is roughly constant.  Of
course, the physical size of the system grows with time, increasing by
panel d, over its initial size, by a factor $\exp(\delta T)=3.7\times
10^7$.  Actually, by this stage, the stream of infalling particles has
ceased and the system, now an isolated one, maintains a fixed size in
$r$ but shrinks in $X$.  This simulation demonstrates the advantage of
scaled variables for studying systems that grow from ``scale-free''
initial conditions.  In the usual physical coordinates, the size and
mass of the system grows by many orders of magnitude whereas the
system is nearly time-independent in scaled variables.  Of course the
variables immediately lose their advantage once the supply of
particles is shut off.

The phase-space plots in Figure 1 are somewhat misleading in that the
shells, each represented by a single point, have masses that range
from $\rho_i\left (R/N\right )^3$ (initially the innermost shell) to
$\rho_i R^2\left (R/N\right )$ (initially the outermost shell).  We
have therefore repeated the experiment but with the mass distributed
equally among the shells.  The results for $T=13.3$ and $18.7$ (final
two frames of Figure 1) are shown in Figure 2.  While these provide
what is at least visually a more accurate representation of the
phase-space distribution function, the simulations used to produce
Figure 1 clearly have better coverage of phase space, and in
particular, are better able to follow the early stages of collapse.

The semi-analytic solutions of FG and B85 are for an eternal
self-similar collapsing system and are described by a semi-infinite
spiral in phase space.  In contrast, our system evolves for a finite
period of time.  Moreover there is an inner region $(r_i<r_c)$ in the
initial perturbation where the density profile deviates from a pure
power-law.  We can see these effects in the phase space plots of
Figure 2 where there appear to be relatively few particles with both
$X$ and $Y$ small.  This region of phase space corresponds to the most
tightly bound particles, i.e., ones with $r_i\to 0$.

FG have shown that the similarity class determines the density profile
of the collapsing system.  For $\delta<1$ ($\epsilon > 2$) the final
density profile is $\rho\propto r^{-\mu}$ where $\mu = 2/\delta$.  For
$\delta > 1$, the final density profile is $\rho\propto r^{-2}$.
Figures 3a, b, and c show the evolution of the density profile with
time for the simulation in Figure 1 as well as two other cases
$\delta\rho_i\propto r^{-2}$ and $\delta\rho_i\propto r^{-3/2}$.  The
agreement with the predictions of FG is quite good. These predictions
have also been confirmed by Moutarde et al.\,(1995) in a rather
different set of numerical experiments.

The density law in the outer regions of the system $(-2\la\log_{10}
X\la -1.2)$ is somewhat steeper than $r^{-\mu}$, a result which is
also evident in the semi-analytic solutions of FG (see their Figure
12).  This region of the system is composed of particles that have
passed through the origin only a few times and is therefore not fully
relaxed.  We see here an example of a system whose evolution is
self-similar but whose density profile is not a pure power-law
function of radius.

As noted above, in N-body simulations, finite resolution effects
introduce an effective short distance cut-off in the perturbation
spectrum as well as the development of small-scale structure.  The
behaviour of the density profile as a function of softening length
$\varepsilon$ (Figure 4) is reminiscent of the results found by Moore
et al.\,(1997).  Like Navarro, Frenk \& White (1996), they find that
the density profile for a dark matter halo is flatter than $r^{-2}$ in
the central regions.  However the size of the region shrinks as one
improves the resolution of the simulations.  Our results provide
further evidence of this effect.

The effects of a softened force law are also evident in the
phase-space plots of Figures 1 and 2 where the maximum velocities
attained by particles as $X\to 0$ approaches a constant.  Were it not
for force softening, this maximum velocity would increase as $X\to 0$

In Figures 5a-d, we show, as a function of logarithmic time $T$, the
``size'' of the system, its scaled mass and energy, and the ratio
$2K/W$.  The system is defined as the region $X<X_{\rm s}$ where
$X_{\rm s}$ is the smallest radius that includes all particles that
have passed through the origin at least once.  These plots show rather
dramatically the speed with which the system reaches the self-similar
state.  At $T\simeq 15$, infall ceases and the system shrinks in
physical variables while remaining more or less stationary in scaled
variables.  Figure 5d is perhaps the most interesting.  We see that
the ratio $2K/W ~ (=2{\cal K}/{\cal W})$ is constant, though different
from $1$, during the infall phase.  The results are consistent with
Eq.\,\EC{virial2} with the terms on the right-hand side being small,
but non-negligible.  Once infall ceases, this ratio settles quickly to
$1$, the value expected for an isolated, virialized system.

\section{Analysis}

\subsection{Self-similar to Virial Transition}

In Section 2 we learned that a system undergoing self-similar infall,
when followed in scaled variables, is ``stationary''.  These results,
together with our numerical simulations, provide clues as to how a
system makes its transition from infall phase to isolated state.  To
the extent that the two terms on the right-hand side of
Eq.\,\EC{virial2} are negligible, this transition is ``smooth'' in
that when the flux of particles through the system boundary is
``turned off'', the system finds itself in a virialized state that is
stationary in the usual sense (i.e., with respect to the physical
variables $r$ and $v_r$).  In this section, we explore the transition
from infall phase to isolated system by attempting to match the SSIM
distribution function to one that describes a system in equilibrium.
The analysis leads to a particular choice for the equilibrium
configuration, namely time-independent self-similar solutions that
correspond to the so-called power-law models discussed elsewhere
(Evans 1994, Henriksen \& Widrow 1995).

Consider a system that is undergoing self-similar infall.  In this
state, the scaled distribution function and potential are independent
of $T$ and therefore satisfy the following  constraint equations
(cf.\,\EC{scale_cbe})

\be{TISCBE}
\left (\delta -1\right ){\cal F} + 
\left (Y-\delta X\right )\frac{\partial{\cal F}}{\partial X}
- \left (\left (\delta -1\right )Y + \frac{\partial{\Psi}} {\partial
X}\right )
\frac{\partial {\cal F}}{\partial Y} = 0
\ee

\noindent and

\be{TISP}
\frac{\partial \Psi}{\partial T} = 0~.
\ee

\noindent If we impose the additional constraint that at the instant 
infall ceases, the system is also time-independent then:

\be{condition2}
\frac{\partial F}{\partial t} = 0~~~~~~;~~~~~~
\frac{\partial \Phi}{\partial t} = 0~.
\ee

\noindent These latter conditions can be written in terms of the scaled 
variables as:

\be{scond2}
\frac{\partial {\cal F}}{\partial T}  +
\left (\delta-1\right ){\cal F} -\delta X
\frac{\partial {\cal F}}{\partial X} 
-\left (\delta -1\right ) Y
\frac{\partial {\cal F}}{\partial Y}=0
\ee

\noindent and
 
\be{poisson2}
\frac{\partial {\Psi}}{\partial T}+
2\left (\delta-1\right )\Psi
-\delta X \frac{\partial\Psi}{\partial X}=0~.
\ee

\noindent Combining Eqs.\,\EC{TISCBE}-\EC{poisson2} leads to the 
following results:

\be{EoneHalf}
{\cal F} = {\cal F}_0 |{\cal E}|^{1/2}~.
\ee

\be{potentials} 
\Psi=\Psi_0 X^{2(\delta-1)/\delta}
\ee

\noindent and

\be{densitys}
{\cal S}=
\frac{2(3\delta-2)(\delta-1)}{\delta^2}\Psi_0 X^{2(\delta-1)/\delta},
\ee

\noindent where ${\cal E}\equiv Y^2/2 + \Psi(X)$.
These equations provide an analytic solution to the Poisson-CBE pair
where the density and potential are power-law functions of the radial
coordinate $X$.  The solutions are valid for $2/3 < \delta < 1$.  (For
$\delta\ge 1$, there is no natural cut-off for the velocity integral
which does not then converge.  For $\delta < 2/3$ the mass within a
finite radius is infinite.)

It is interesting to note that the physical distribution function has
the same functional form as in Eq.\,\EC{EoneHalf}: $F=F_0 |E|^{1/2}$
where $E\equiv v_r^2/2+\Phi(r)$ is the usual particle energy.  The
fact that $F$ depends on $v_r$ and $r$ only through $E$ follows
from the condition that at the instant infall ceases, the system is
time-independent (the Jeans theorem).  The specific form found, the
$E^{1/2}-law$, is a result of the constraint that the system be
self-similar during the infall phase and continuously into the
virialized state.

Our conjecture is that Eqs.\,\EC{EoneHalf} and \EC{potentials}
describe the system up to the instant when infall ceases and that
after this, the distribution function is given by $F(E)=F_0
|E|^{1/2}$.  This conjecture is subject to the following caveats: 1)
The solution is reached asymptotically and in general will apply to
the most tightly bound particles.  2) The system must match onto the
cosmological background beyond its boundary and therefore
the potential and density must change near $X_{\rm s}$; 3) The system
does undergo some readjustment after infall cease, e.g., $2K/W$ does
change, though not by very much.

Eqs.\,\EC{potentials} and \EC{densitys} might, by continuity, be taken
to imply that during the infall phase the physical potential and
density, $\Phi$ and $\rho$, are time-independent.  However, during the
infall phase, the system itself is not time-independent: Quite to the
contrary, the size of the relaxed region is increasing as $t^\delta$.
The new material that falls in does not change the density profile in
the already-relaxed region and so in this sense the system is steady.
Thus it is not surprising that the $E^{1/2}-{\it solutions}$ have been
encountered in studies of equilibrium solutions of the CBE (Evans
1994, Henriksen \& Widrow 1995).  In Henriksen \& Widrow (1995), we
surveyed all self-similar steady-state solutions with radial orbits in
spherical symmetry.  These were found to be the very same power-law
solutions found here (cf. Eqs.(2.20) and (2.22) in that paper with
Eqs.\,\EC{potentials} and \EC{densitys}).  Note that $\delta_{{\rm
HW}95}=\delta/(\delta-1)$ so that the allowed range for $\delta$ found
here corresponds to $\delta_{{\rm HW}95}< -2$.  The $E^{1/2}-law$ for
the distribution function was also discussed there.  Indeed, one can
show, using the standard solution to Abel's integral equation (e.g.,
Binney and Tremaine, 1987) that {\it any} power-law density profile
for a system of particles on radial orbits leads to this form for the
distribution function.

Figure 6 gives the distribution function ${\cal F}({\cal E})$ for our
simulated halo.  We have chosen a time-frame close to the point where
the last particles are falling in.  As expected, the agreement with
the $E^{1/2}-{\it law}$ is best for the most tightly bound particles.

\subsection{Negative Temperature Models for the Distribution Function}

It is of fundamental interest in astrophysics and cosmology to
understand complicated systems like galaxies and clusters in terms of
simple distribution functions.  For an isolated system in equilibrium,
the distribution function will depend only on the integrals of motion.
Merritt, Tremaine, and Johnstone (1989) have investigated models
based on the following functional form for the distribution function:

\be{stbt}
f = f(E,j^2) = 
\left \{ \begin{array}{ll}
A\left (-E\right )^{3/2} e^{a\left (E+j^2/2r_a^2\right )}
 & \mbox{$E<0$} \\
0                                & \mbox{$E\ge 0$}
\end{array}
\right.
\ee

\noindent where ${A, a, r_a}$ are positive parameters.  $a>0$ corresponds
to ``negative temperature'' and represents the essential distinction
between these models and those considered by Stiavelli and Bertin
(1986).  Figure 7 shows this distribution function in $(r,v_r)$ phase
space.  We note that the effect of the exponential is primarily to
``empty'' the central region of phase space corresponding to what
would be the most tightly bound particles.  It is not clear whether
this is a physically plausible equilibrium configuration in light of
the radial instability discovered by He\'non (1973) (see also Barnes,
Goodman, \& Hut 1986).

Eq.\,\EC{stbt} was arrived by considering a crude model of violent
relaxation (Lynden-Bell 1967).  In particular, the energy dependence
assumes that the relaxation process redistributes the energies of
particles as they pass through regions of rapidly varying
gravitational potential (Tremaine 1987; Merritt, Tremaine, and
Johnstone 1989).  However, the process of structure formation in
hierarchical clustering models may proceed through relaxation that is
violent enough to wash out the intricate phase space structure of the
FG and B85 solutions but not so violent as to remove all memory of the
initial particle ordering.  To illustrate this last point, we plot
initial vs. final energies for the particles in the simulation (Figure
8).  While there is a certain amount of scatter due to the
instability, the correlation is clear: particles remember where they
were in the initial perturbation.  This point was discussed in the
context of full 3-dimensional N-body simulations by Quinn and Zurek
(1988).

We propose an alternative distribution function, motivated by the
results of the previous section, and found by replacing the
$(-E)^{3/2}$ factor in Eq.\,\EC{stbt} with $(-E)^{1/2}$.  We restrict
our discussion to strictly radial orbits and consider a distribution
function of the form

\be{alt_stbt}
f({\bf r},\,{\bf v}) = \left \{ \begin{array}{ll}
A\left (-E\right )^{1/2} e^{aE}\delta\left (j^2\right ) & \mbox{$E<0$} \\
0                                & \mbox{$E\ge 0$}
\end{array}
\right.
\ee

\noindent The analysis of this family of models is straightforward 
and the details are left for the appendix.  The distribution function
in $(r,v_r)$ phase space is shown in Figure 9.  Note that the
distribution function does decrease as $E\to -\infty$ though not as
dramatically as in the model described by Eq.\,\EC{stbt}.  Our
conjecture is that Eq.\,\EC{alt_stbt} also describes, at least
approximately, the distribution function during the self-similar
phase, provided we substitute ${\cal E}$ for $E$.  This would allow
for arguments similar to those found in the previous section where
the system finds itself in an equilibrium configuration once infall
ceases.

As noted above, the effect of the exponential is to ``empty'' the
central region of phase space.  Physically, this might arise because
the initial density profile is not strictly a power-law function of
radius.  In Eq.\,\EC{pert_def}, for example, the density as $r\to 0$
is less than what it would be for a pure power-law profile.  This
manifests itself in slightly fewer particles in the central regions of
phase space in the final collapsed object, as can be seen in Figure 2.
In addition, the mild form of violent relaxation that occurs in our
model may also lead to a cut-off at large negative energies in the
distribution function.

Given an empty center in phase space, one can show, by methods 
similar to those in the appendix, that $\Psi(X)\propto -\left (
-\log(X)\right )^{2/3}$ and thus $\rho t^2\propto 
\left (-\log{X}\right)^{-1/3} X^{-2}$.  Our proposed form
for the distribution function, Eq.\,\EC{alt_stbt}, requires only
a weak logarithmic correction to the inverse square law (the
time dependence depends on $\delta$).

\subsection{$\delta=1$: Boundary of Self-similar Behaviour}
 
To complete our discussion of the SSIM we turn to initially flat
profiles ($\delta>1; \epsilon<2$) and to the transition case
($\delta=1; \epsilon = 2$). It is clear, both from our simulations and
from the work of FG, that there is no strictly self-similar evolution
for these initial conditions.  This can be understood in a variety of
ways.  Consider first the phase diagram for the characteristic curves
of the self-similar CBE.  It is useful to define an effective
potential, $\Psi_{\rm eff}\equiv\Psi + \delta(\delta-1)X^2/2$ (see
HW).  For $\delta<1$ (Figure 2 of HW) there is a saddle point
singularity at $Y=\delta X$ and $d\Psi/dX_{\rm eff} = 0$, i.e.,
simultaneously a turning point and an extremum in the effective
potential.  However, when $\delta>1$, the effective potential
increases monotonically and there are no singular points at finite
$X$.  Since it is the appropriate pair of separatrices of the saddle
point that form an effective boundary to the relaxed region of phase
space, the $\delta>1$ case corresponds to the situation where the
region of relaxation extends to infinity.  Infalling particles
continuously perturb the inner regions of the halo.  This last comment
can be understood from the observation that the potential per unit
mass increases outward for $\delta>1$.  The implication is that in
this case the system is $T$-dependent in our scaled variables. Our
simulations as well as others (e.g. Moutarde et al., 1995) show
however that the system evolves toward the $r^{-2}$ profile, the
boundary of the so-called steep cases.

The $\delta=1$ case has $X=r/t$ and is an example of the general class
of self-similarity called `homothetic' (e.g.\,Carter and Henriksen,
1991).  It has been studied in the context of isothermal Newtonian
collapse and in spherically-symmetric General Relativistic collapse.
It is recognized here, in the context of the CBE-Poisson system, for
the first time.

Following HW we consider the characteristic curves for the 
CBE-Poisson system.  Along a given characteristic,

\be{chardelta1}
\frac{d Y}{d X}~=~\frac{1}{X-Y}\frac{d\Psi}{dX}
\ee

\noindent and
 
\be{dfonchar} 
{\cal F}~=~{\rm constant}
\ee

\noindent (cf. Eq.\,\EC{scale_cbe} and \EC{scale_poisson} with
$\delta=1$ and $\partial/\partial T = 0$).  It is a simplifying property of 
this case that ${\cal F}$ is constant on a characteristic.

We label characteristic curves by the continuous variable $\zeta$ so
that ${\cal F}={\cal F}(\zeta)$ and $Y=Y(X,\zeta)$.  The right-hand
side of the Poisson equation can then be transformed to an integral
over $\zeta$:

\be{characterp} 
\frac{d}{dX}\left(X^2\frac{d\Psi}{dX}\right)=\int~{\cal F}(\zeta)\,\frac{
\partial Y}{\partial\zeta}\,d\zeta~.
\ee

\noindent Eqs.\,\EC{chardelta1} and \EC{characterp} pose a rather 
intractable integro-differential problem.  Fortunately the
self-similar nature of this case suggests an approximate form for the
density profile and potential (i.e., $\rho\propto r^{-2}$ and
$\Psi=\Psi_o\ln{X}$) which can be used as an initial guess in an
iterative attack on the problem.  With this choice for the potential,
the equation for a characteristic curve reduces to

\be{first_interate}
\frac{dY}{dX}=\frac{\Psi_o}{X(X-Y)}. 
\ee

\noindent This equation may be integrated to give the family of 
characteristics implicitly:

\be{soln_character}
\frac{1}{x}=e^{y^2}\left(\frac{1}{\zeta}-{\rm erf}(y)\right),
\ee
 
\noindent where $X\equiv x\sqrt{2\Psi_o\pi}$ and 
$Y\equiv y\sqrt{2\Psi_o}$.  The iteration procedure proceeds by taking
the above solution for $Y(X,\zeta)$ and using it to perform the
integration in the Poisson equation.

Examples from this family of curves are shown in Figure 10.  This
figure is reminiscent of the FG and B85 solutions.  However each of
those solutions represents a single-stream trajectory in phase space.
Conversely, each curve in Figure 10 is regarded as an independent
characteristic in the solution to the CB and Poisson equations, at
least to this order of approximation.  This represents an essential
departure from the FG and B85 approach wherein each `loop' is
populated sequentially from the same stream of particles so that
${\cal F}$ is non-zero along a one-dimensional curve in phase
space. In our approach phase space is populated not only by this
orderly `phase mixing', but also through the HW instability. This
allows the phase-space density to take up its most general variation
as allowed by the characteristic curves of the CBE. This is perhaps
the best definition of a `relaxed' system.
   
Still this last argument does not explain why the system with a flat
initial profile is driven toward the self-similar ``boundary'' rather
than away from it.  This question is of course very general and is not
unrelated to the question as to why a violently relaxing, isolated,
collisionless system eventually virializes.  Some clues are provided
by Tremaine, H\'enon and Lynden-Bell (1986) and references therein.
In particular they show that equilibrium is always attained as the
maximum of an H-function.  The H-function is some (unfortunately
undefined) integral over a functional of the distribution function.

We propose that there exists a similar H-theorem that describes the
onset of self-similarity.  It is, however, necessary to consider a
more general integrands than those considered by Tremaine, H\'enon and
Lynden-Bell (1986).  The extension to more general integrands is by
way of the Lyapounov functional idea (Berryman, 1980). Essentially in
the present context, this reduces to finding a functional of the form

\begin{equation} 
{\cal H}=-\int ~ h({\cal F}, X,Y)\,dX\,dY,
\end{equation}

\noindent that increases throughout the collisionless evolution, and 
that is a finite constant in the self-similar state. Because the
transformation to scaled variables leads to a transformation of the
physical quantities by various powers of $e^T$, we must in general
require that H vary more rapidly than an exponential, i.e., that it
reach the self-similar state in a finite time. This last requirement
precludes candidates like $-{\cal M}$ which otherwise have the properties
of a Lyapounov functional.  Such a choice would, in fact, give no
information since the logarithmic derivative of the functional is then
constant and there is no real approach to the self- similar state.

One possible functional is 

\be{functional} 
{\cal H}=-\frac{1}{2}\int ~{\cal F}\,(Y-\delta X)^2\,dX\,dY,
\ee

\noindent provided that the integral is restricted to the
region of phase space for which $(Y-\delta X)(\partial {\Psi}_{\rm
eff}/\partial X) >0$.  This last condition essentially restricts the
integration to the region of phase space where $v_r>\delta r/t$ since
$\partial\Psi_{\rm eff} /\partial X >0$, at least for the cosmological
cases considered here.  The condition also means that particles that
are still taking part in the expansion (i.e., those particles at large
$X$) are excluded from the integration since these have $v_r = 2r/3t$.
$\cal H$ is clearly constant in the self-similar state since
$\partial_T {\cal F}=0$.  However, time-dependence in scaled variables
(as well as physical variables) arises as the
``spiral pattern'' of the FG and B85 solution is filled in.  This
implies that the corresponding function in physical variables,
$H\equiv {\cal H}\exp{(5\delta-4)T}$ is {\it not} a constant (which
would render ${\cal H}$ trivial) so long as particles are falling in.
One can show explicitly that $\partial_T{\cal H}>0$ since infall
increases the number of outward going particles and $\delta r/t$
decreases with time.  We see that the system is driven to the
self-similar state by the continuing infall of new particles.  Once the
source of particles is removed, self-similar infall gives way to
self-similar equilibrium.

It thus seems possible that generalized Lyapounov functionals exist
whose monotonic increase measure the approach to self-similarity just
as the H-functional of Tremaine, H\'enon and Lynden-Bell (1986) for
the approach to equilibrium.  In fact similar arguments may be carried
out for the steep profiles ($\delta <1$) using an integrand
$h=\frac{1}{2} {\cal F} X^2$ and carrying the integration over the
inward going particles. In this case the system remembers the
self-similarity in the ultimately steady state.

\subsection{Keplerian Halos}

A second example that can be treated analytically is that of test
particles on radial orbits outside a fixed mass.  Such a situation
arises when accretion continues after the bulk of the system
mass has fallen in.  We expect that this `Keplerian' set of particles
will pursue a self-similar evolution with similarity class $\delta
=2/3$ since now the effective remembered constant is $GM$ where $M$ is
the mass that accumulates during the primary accretion phase (see e.g.
Henriksen \& Widrow 1995).  This reveals immediately the possible
interest since now the power law part of the self-similar evolution
will have a density profile $\rho\propto r^{-2/\delta}$ or $r^{-3}$
and hence might explain the outer steep part of the NFW profile.
 
The characteristics of the CBE delineate the shape of the eventual
relaxed region in phase space.  The topology of the phase diagram is
typical of steep self-similar evolution.  The equation of the
characteristics in scaled variables is found from Eq.\,(8) of HW by
setting $\delta=2/3$ and $\Phi=\Psi=-M/X$.  The family of curves has
the anticipated saddle point at $X_c^3=9{\cal M}/2$ and
$Y_c=2X_c/3$. The discussion is simplest in terms of the variables
$u\equiv (Y-X/3)/(X_c/3)$, and $x\equiv X/X_c$.  Then the family of
characteristics is given by

\be{twothirds1}
\frac{du}{dx}=\frac{2(x-1/x^2)}{(u-x)}~.  
\ee
  
A sample of characteristic curves is shown in Figure 11.  Figure 11a
is constructed in terms of the variables $u$ and $x$.  The critical
curves are not drawn explicitly, but it is relatively easy to identify
their slopes ($1/3$ and $-1/2$) as the limiting curves passing through
$(1,1)$ (the `x' in the diagram). These limiting curves connect
respectively the stable nodes (lower left to upper right) and unstable
nodes (upper left to lower right) that exist at infinity.  Figure 11b
shows the same curves, but drawn in terms of the variables used
in other plots in the paper.
  
We expect that the only curves to be populated are those in the left
part of the figure (gravitationally bound particles) together with an
accretion stream which joins the node at the upper right.  There is of
course an inner boundary for these types of particles corresponding to
the outer edge of the main system.  Roughly speaking, the radius of 
the transition region will correspond to the parameter $R_s$ in the
NFW profile.
  
\section{Summary and Conclusions}

N-body simulations of collisionless, self-gravitating matter, such as
those by Navarro, Frenk, and White (1996) and Moore et al.\,(1997),
have provided some tantalizing results, in particular suggesting that
nonlinear structure in the Universe possesses certain scale-invariant
or self-similar characteristics.  These results are perhaps not
surprising given that most simulations assume an Einstein-de Sitter
Universe and an at least approximately scale-free initial spectrum of
density perturbations.  With these assumptions, there is only one
characteristic scale in the Universe, the expansion rate.  This
observation has been exploited by Press and Schecter (1974), Bower
(1991), Lacey and Cole (1993), and others to formulate an analytic
model for the evolution of nonlinear structure within the framework of
the hierarchical clustering scenario.  However, these models are
limited in scope.  While they provide predictions for distribution of
collapsed objects as a function of mass, they say little about the
dynamical process by which a system relaxes to a virialized (or
quasi-virialized) state.

The SSIM is, in some respects, complementary to a Press-Schecter type
formalism.  The model begins with highly idealized initial conditions
(strictly power-law initial density profile; no angular momentum) but
allows one to follow in detail the complete evolution of a system.
Analytic calculations and numerical simulations provide clues
as to the density profile and distribution function of the system 
both during the infall phase and in the final equilibrium state.

Our results can be summarized as follows:

\begin{itemize}

\item  Soon after gravitational collapse begins, the system,
driven by infall of mass through its boundary, enters a period of
self-similar evolution.  The system quickly virializes once infall has
ceased.  It is likely that the final virialized state is one of the 
stationary self-similar family discussed by Evans (1994) and
Henriksen \& Widrow (1995) with the self-similar `class' $\delta$
remembered from the dynamic self-similar infall phase.

\item  During the infall phase, the system relaxes through
a combination of phase mixing, phase space instability, and {\it
moderate} violent relaxation.  However, relaxation does {\it not}
completely randomize particle energies, i.e., particles maintain some
memory of their initial state.

\item  As the system relaxes, the single trajectory of the FG and 
B85-type solution is transformed into a continuous distribution in
phase space.  The subsequent cycles of the single trajectory are now
regarded as characteristic curves of the smooth distribution function.
We have used this approach to study two similarity classes that are
important to our arguments; $\delta=1,2/3$.

\item  The self-similar phase may be recognized as ``stationary''
in the appropriate scaled variables.  By following the development of
a perturbation in these variables, we can achieve a remarkable dynamic
range in our simulations.  

The size of the system is always a fixed fraction of the turn-around
radius.  In addition, the relation for the mass of the system as a
function of time is in agreement with that found by Lacey and Cole
(1993).

\item  During the self-similar phase, the system obeys a virial condition 
$2K/W={\it constant}$.  The constant differs from the usual value $1$
(though by only about 10\%) due to mass flux through the system
boundary and the time-dependent nature of the infall solution.

\item  From the observation that the transition from infall phase to
isolated state is relatively gentle, follows the conclusion that the
distribution function for the final object has the form $F\propto
\left (-E\right )^{1/2}$.  In a realistic system, we expect that 
the distribution function will eventually decrease at large negative
energy and so we include an exponential negative temperature factor.
This leads to a form for the distribution function that is similar to
the ones proposed by Stiavelli \& Bertin (1985) and Merritt, Tremaine,
and Johnstone (1989) and reminiscent of the Fermi-type function
proposed by Lynden-Bell (1967).

\item  The SSIM predicts an effectively universal profile
($\rho\propto r^{-\mu}~{\rm with}~\mu\simeq 2$) for the intermediate
region of a dark matter halo.  The system does remember the initial
profile for $\epsilon>2$ (the so-called steep cases) but the effects
on the final density profile are relatively minor.  For flat initial
profiles, the system is driven towards the limiting self-similar
profile ($\delta=1$) by accretion of spherical shells with
ever-increasing binding energy, a process reminiscent of that
considered by Syer and White (1997).  In this sense, the density
profile is a one-sided attractor.

Because of finite lifetime effects, the inevitable breaking of scale
invariance at the centre of the initial perturbation, and the
presence of angular momentum, one can expect a physical (although
ill-defined) flattening near the centre of the relaxed halo.  In
addition, the simulations themselves are sensitive to resolution
effects in this region and so the actual flattening may be less
pronounced than in the NFW profile.

\item  The outer parts of the halo, still in a self-similar 
relaxation phase, can be quite steep either because the particles
there are not full relaxed or because they are essentially in 
Keplerian orbits about the bulk of the halo mass.  In this latter
case, the power law is $r^{-3}$ in accordance with the NFW profile
and corresponding to the similarity class $\delta=2/3$.

\item In our simulations, the outer edge of the initial mass distribution,
$r_0$, determines ultimately the radius of the transition region between
$r^{-2}$ and $r^{-3}$ behaviour ($R_{\rm s}$ in the NFW profile).
Likewise, the total mass in the simulation fixes $M_{\rm s}$.

\item  The evolution of the initially flat systems towards a
limiting self-similar state during the infall phase recalls the
evolution of an isolated thermodynamic system towards a maximum
entropy state.  Here we propose Lyapounov functions that are maximized
in the self-similar state and which are monotonic under collisionless
evolution of the distribution function during continued infall.  The
existence of such functions tends to strengthen our belief in the
`universality' of the SSIM.

\end{itemize}

\vskip 1in

\centerline{\bf Acknowledgements}

This work was supported in part by a grant from the Natural Sciences
and Engineering Research Council of Canada.

\newpage

\appendix{\bf Appendix}

In this appendix we discuss equilibrium models for spherically
symmetric systems of collisionless particles on completely radial
orbits.  In particular, we consider the following functional form for
the distribution function:

\be{a1}
f({\bf r},\,{\bf v}) = \left \{ \begin{array}{ll}
A\left (-E\right )^\alpha e^{aE}\delta\left (j^2\right ) & \mbox{$E<0$} \\
0                                & \mbox{$E\ge 0$}
\end{array}
\right.
\ee

\noindent where $A$, $a$, and $\alpha$ are positive constants.
In the negative temperature models of Merritt, Tremaine, and Johnstone
(1989), $\alpha=3/2$.  (The precursor to these models, studied by
Stiavelli and Bertin (1985), also had $\alpha=3/2$ but with $a<0$,
i.e., ``positive temperature).  On the other hand, our analysis
suggests $\alpha=1/2$, at least in the relaxed region.

We begin with the Poisson equation.  Following Merritt, Tremaine, and
Johnstone (1989), we define the auxiliary potential, $W=-a\Phi$ and 
change integration variables from $v_r$ to $x$ where $x^2\equiv
-aE = W - av_r^2$.   The Poisson equation then becomes:

\be{a2}
\frac{d}{dr}r^2\frac{d}{dr}W = -\frac{g(W)}{\gamma}
\ee

\noindent where $\gamma\equiv a^{n-1/2}/2\pi GA$ and

\be{a3}
g(W)\equiv 2^{5/2}\pi\int_0^{\sqrt{W}}
\frac{dx x^{2\alpha+1} e^{-x^2}}{\sqrt{W-x^2}}
\ee

\noindent At this stage, Merritt, Tremaine, Johnstone (1989)
introduce the `homology invariant functions' $z=W$ and $y=-r dW/dr$
(Chandrasekhar 1939) thereby reducing the Poisson equation to
a first order ODE.  This technique does not seem appropriate
for $\alpha=1/2$ and we choose instead to solve Eq.\,\EC{a2}
directly.  

It is convenient to change variables from $r$ to $s\equiv -\ln{(r)}$,

\be{a4}
\frac{d^2 W}{ds^2}-\frac{dW}{ds}=-\frac{g(W)}{\gamma}~,
\ee

\noindent and integrate from large $r$ toward the origin, i.e., $s=s_i$
to $s=s_f$ where $s_i<0$ (e.g., $s_i=-15$) and $s_f>0$ (e.g., $s_f=15$).
At large $r$, $W\ll 0$ and $g(W)= \sqrt{2}\pi^2 W/\gamma$.  
Asymptotic solutions are easily found:

\be{a5}
W = W_+ e^{\beta_+ s} + W_- e^{\beta_- s}
\ee

\noindent where $\beta_\pm = \frac{1}{2}\left (1\pm\sqrt{1-4\sqrt{2}
\pi^2/\gamma}\right )$.  At large $r$ ($s\to -\infty$) the
$\beta_-$ solution dominates and corresponds to power-law behaviour
for the potential: $\Phi\propto r^{-\beta_-}$.  In general, initial
conditions at finite $s$ must include an admixture of both solutions
in order to satisfy the appropriate boundary conditions at the origin,
namely that the mass be positive definite but not include a point mass
at $r=0$.  

We use standard numerical techniques to find the potential.  With the
potential in hand, the distribution function, $F(r,v_r)$ is easily
computed leading the gray-scale plots in Figure 7 and 9.

\newpage

\centerline{\bf Figure Captions}

{\bf Figure 1:} Evolution of a system of spherical shells in
$(X,Y)$-phase space.  The initial density perturbation is a power law
function of radius $\delta\rho_i\propto r^{-2.5}$ except near the
origin (see text).  Initial velocities follow the unperturbed Hubble
flow.  The panels a-d correspond to logarithmic times $T=2.7,\, 8.0,\,
13.3$ and $18.7$ respectively.

{\bf Figure 2:} Same as Figure 1 (only $T=13.3$ and $18.7$ are shown)
but with equal mass particles.

{\bf Figure 3:} Density as a function of $X$ for 5 different choices
of $T$ with $T$ increasing from bottom to top.  We plot $\rho
r^3\propto X{\cal S}\exp{(3\delta -2)T}$.  (If instead we were to plot
$X{\cal S}$ the curves would lie on top of one another.)  The initial
density profiles are given by $\delta\rho_i\propto r^{-\epsilon}$ with
$\epsilon = 2.5,\, 2.0,$ and $1.5$ for Figures 3a, 3b, and 3c
respectively.  Dashed lines are the predictions from FG, for
$\epsilon=2.5$, $X{\cal S}\propto X^{6/7}$ and for $\epsilon=2.0$ and
$1.5$, $X{\cal S}\propto r^{1}$.

{\bf Figure 4:} Density law for different values of the softening
length $\varepsilon$ at two different times.  The solid curves are
for $\varepsilon = 0.0005$; dotted curves are for $\varepsilon
= 0.0025$; dashed curves are for $\varepsilon = 0.01$. 

{\bf Figure 5:} System-wide quantities as a function of $T$.  Note
that infall ceases at $T=15$.  Panel a gives the scaled size of the
system, $X_{\rm s}$, defined in the text.  Panels b and c give the
total scaled mass and energy within $X_{\rm s}$.  Panel d gives the
virial ratio $2K/W = 2{\cal K}/{\cal W}$.

{\bf Figure 6:} $F(E)$ vs E: the
distribution of particles as a function of energy,

{\bf Figure 7:} Gray-scale plots of the distribution function in
$r, v_r$ phase space for the so-called negative temperature
distribution function of Merritt, Tremaine, and Johnstone (1989).

{\bf Figure 8:} Scatter plot of initial vs. final particle energies
for the same simulation as shown in Figure 1

{\bf Figure 9} Same as Figure 7 but for the $|E|^{1/2} \exp{(aE)}$
distribution function suggested in this work.

{\bf Figure 10:} Semi-analytic solutions for the characteristic curves
of the CB and Poisson system with $\delta=1$.

{\bf Figure 11:} Same as Figure 10 but for $\delta=2/3$.

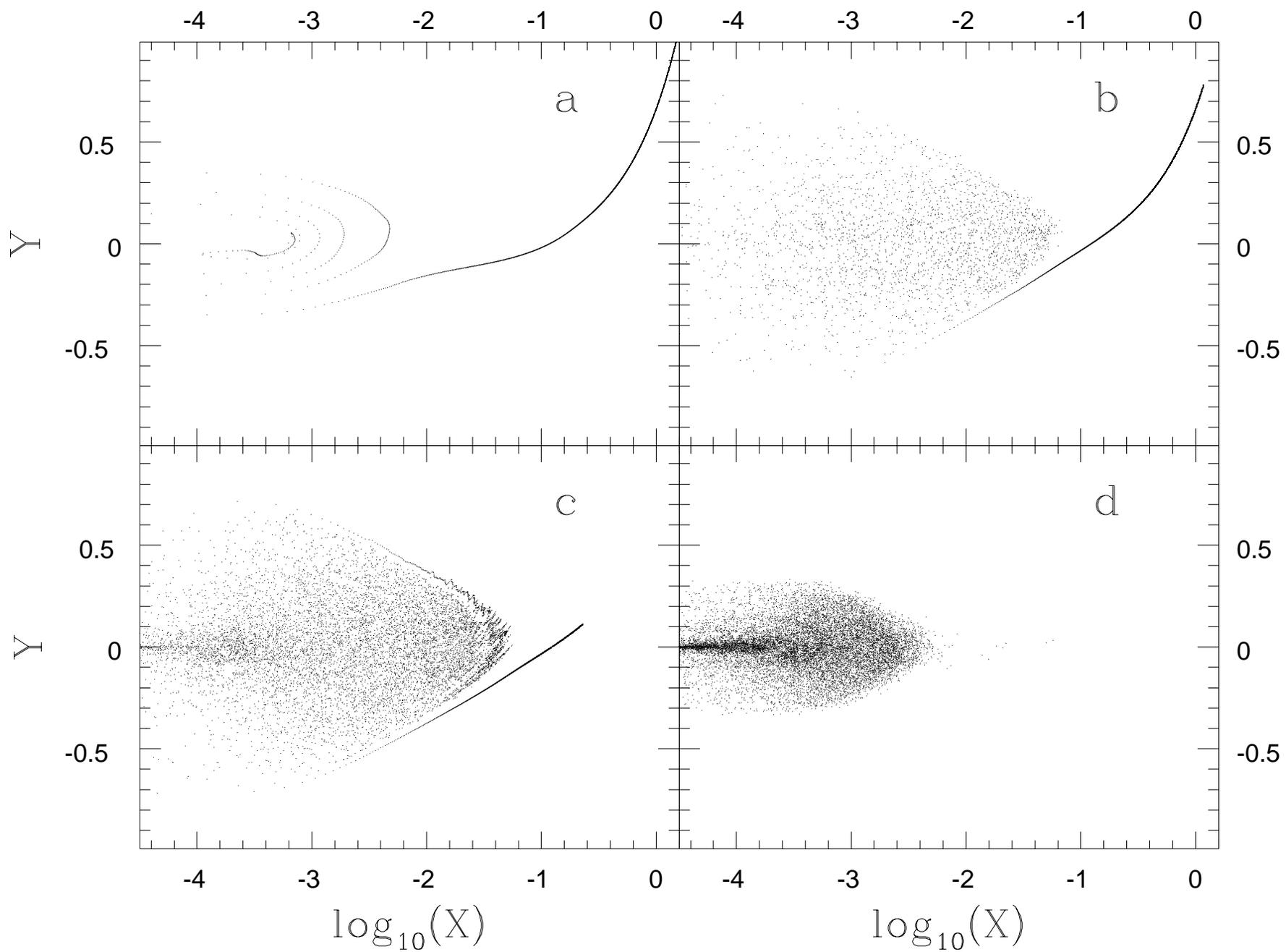
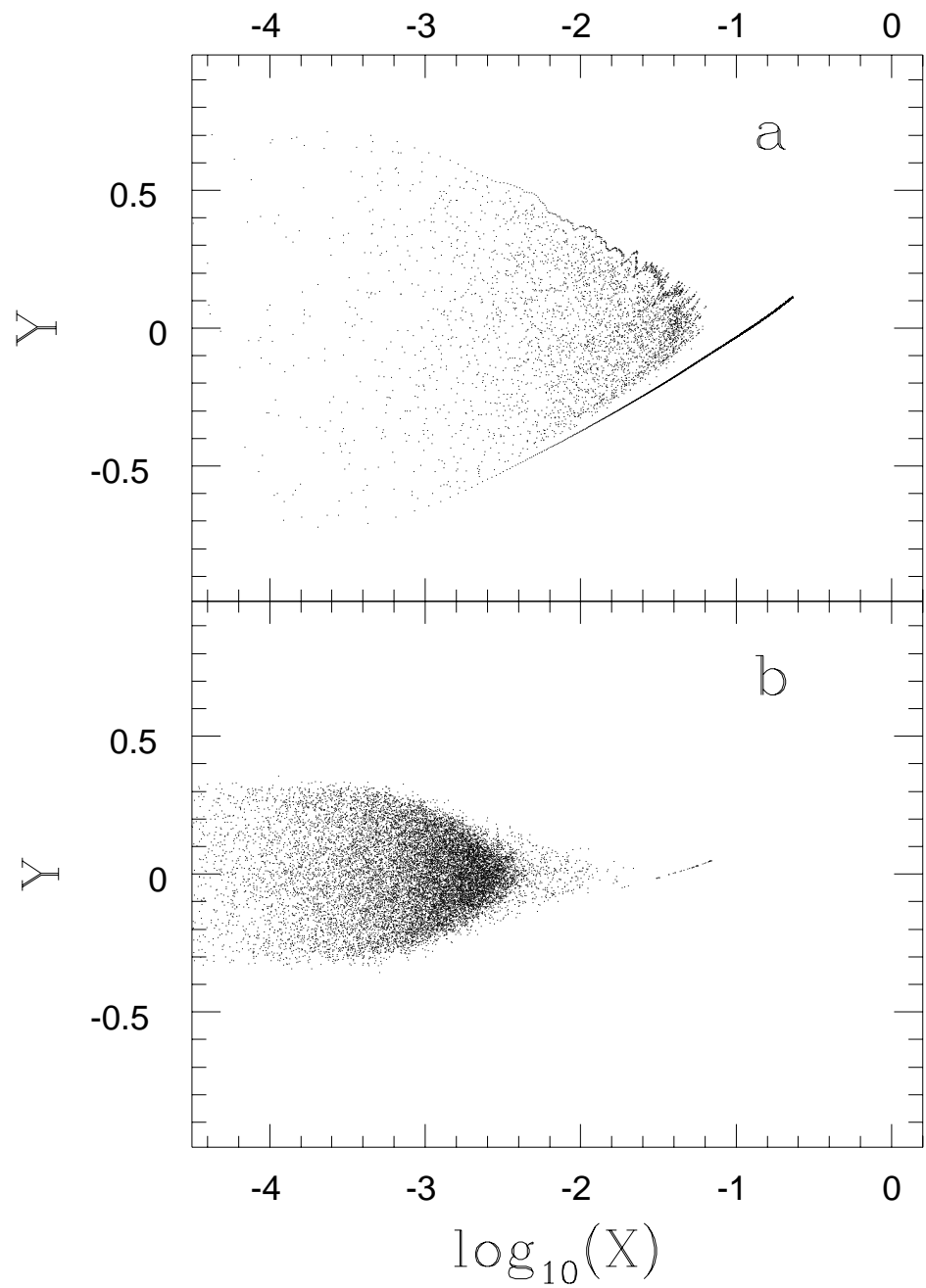
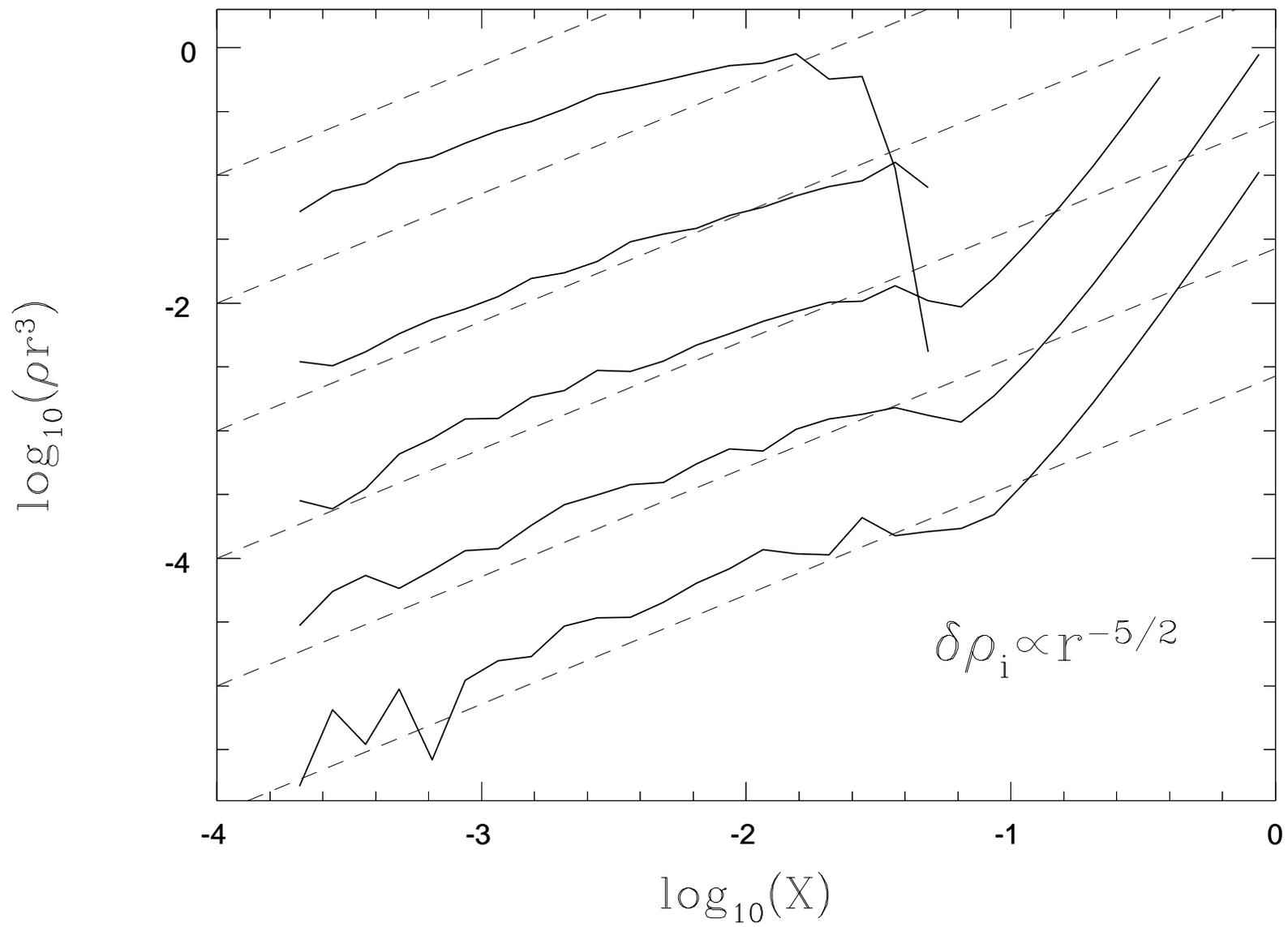
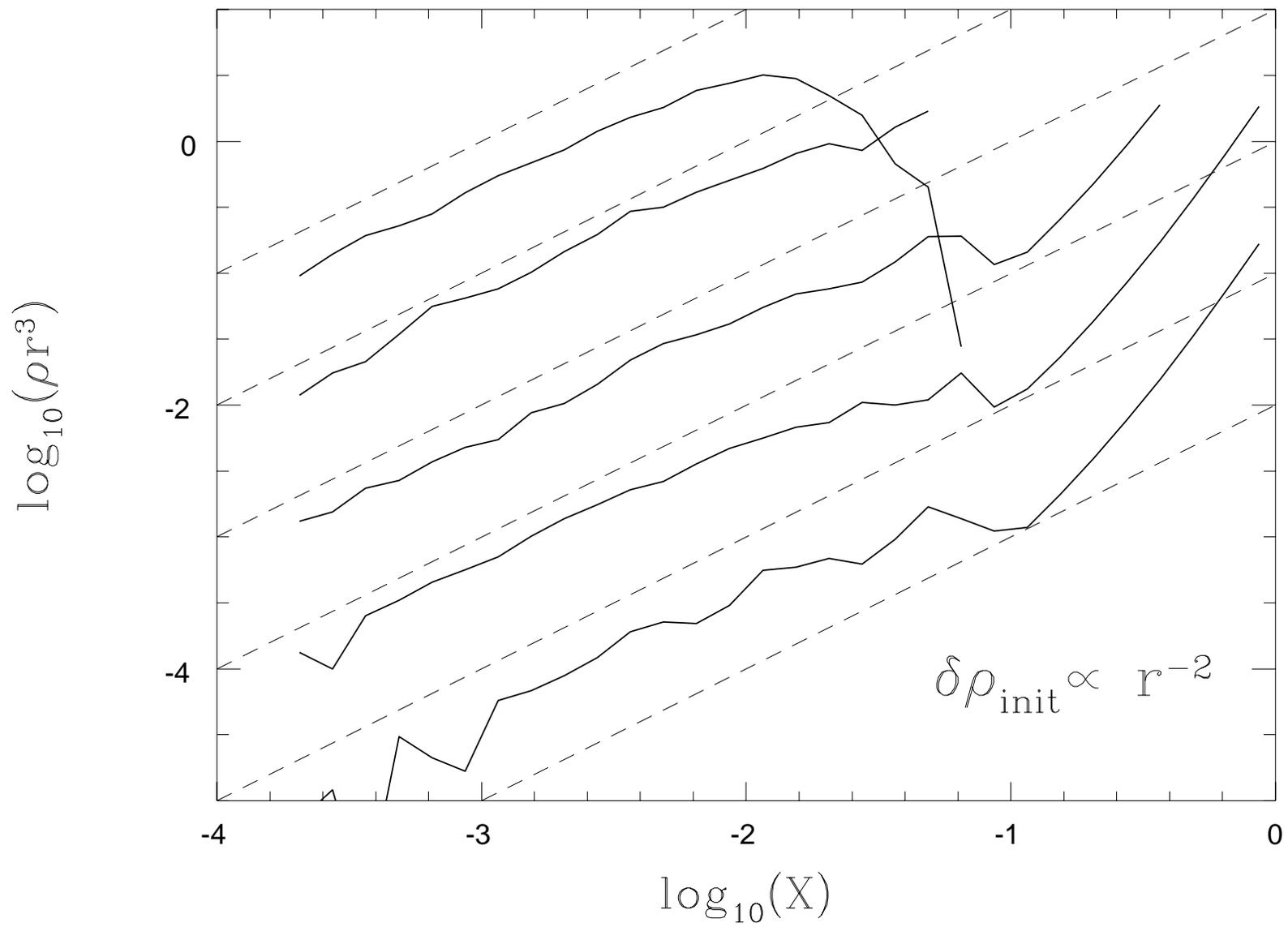
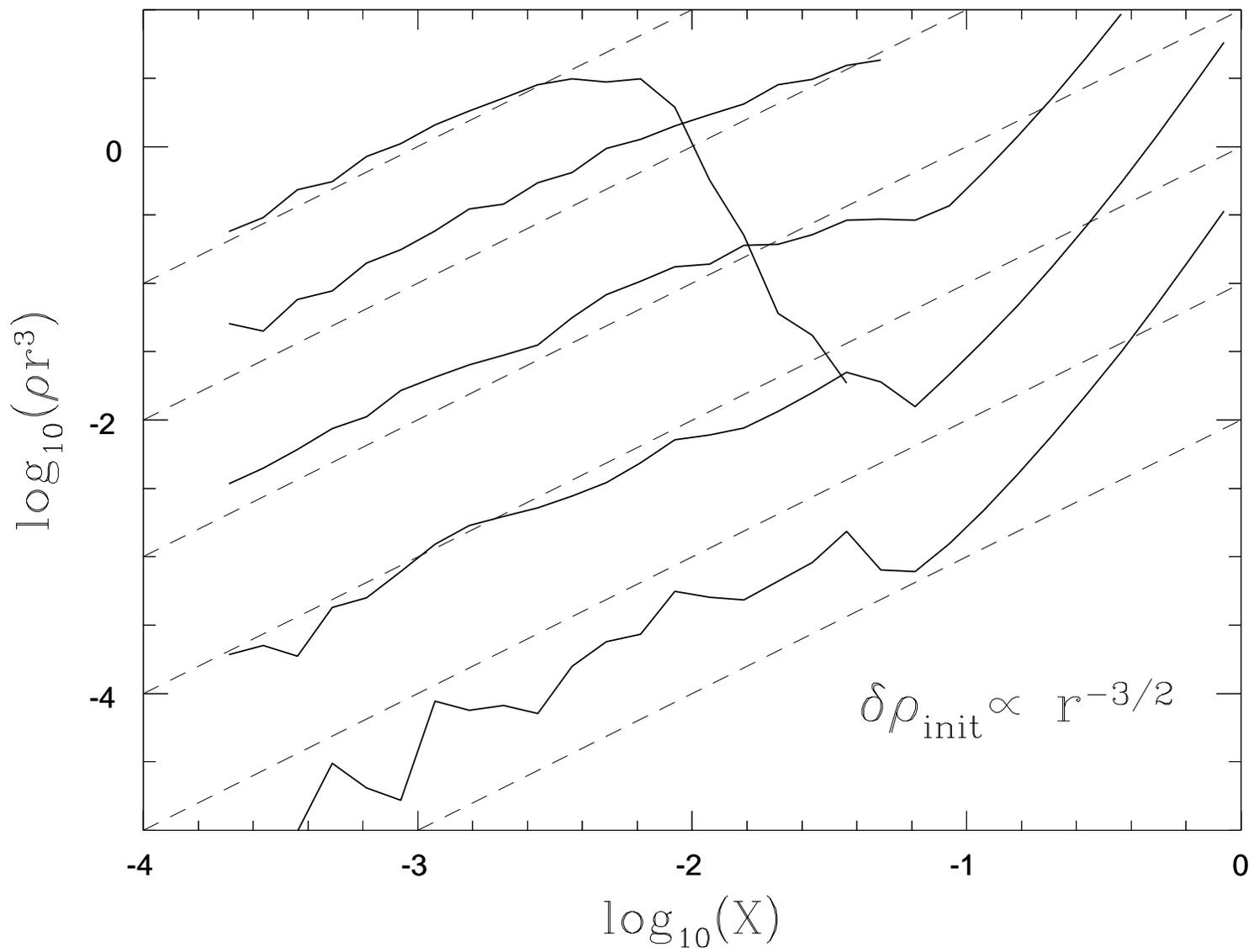
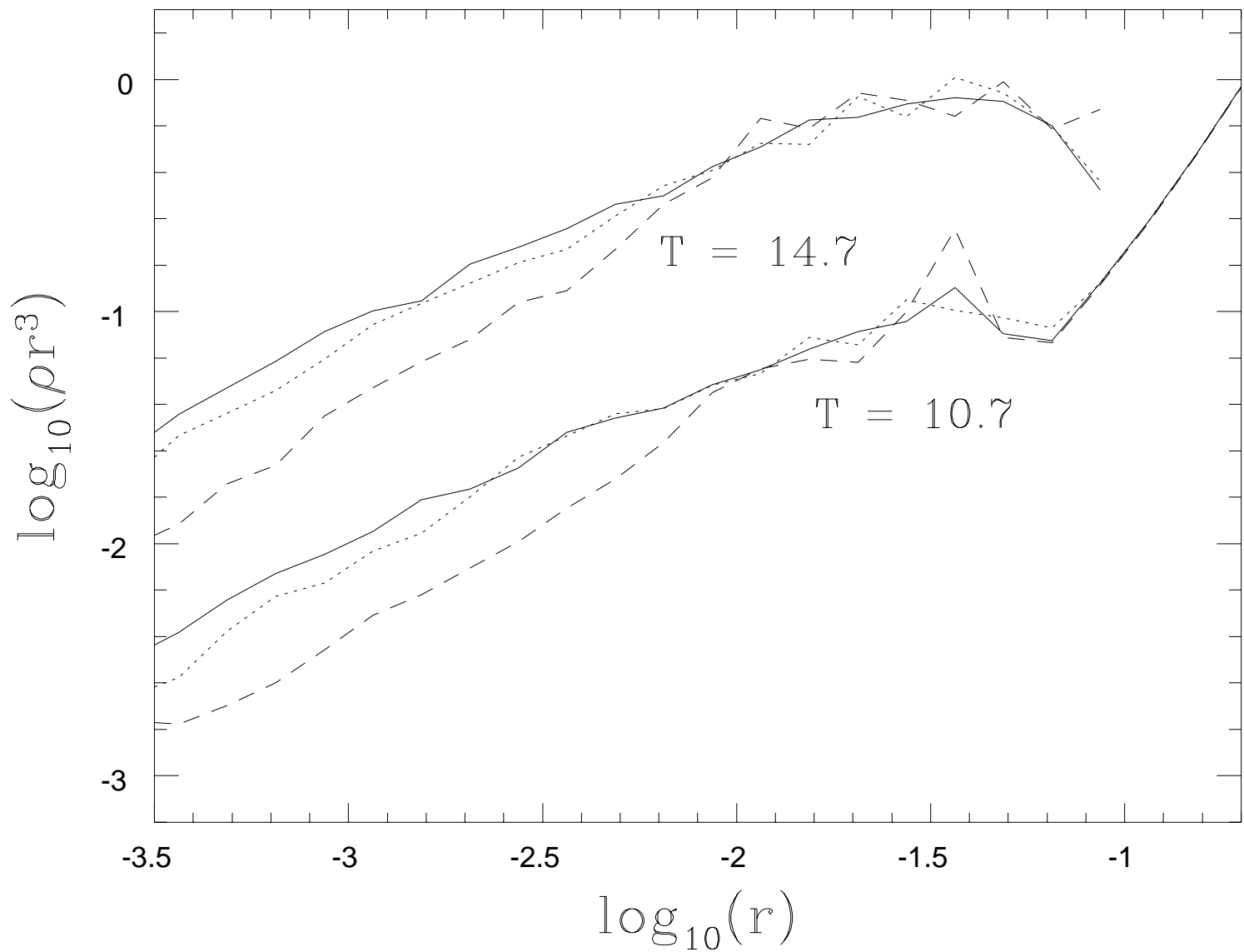
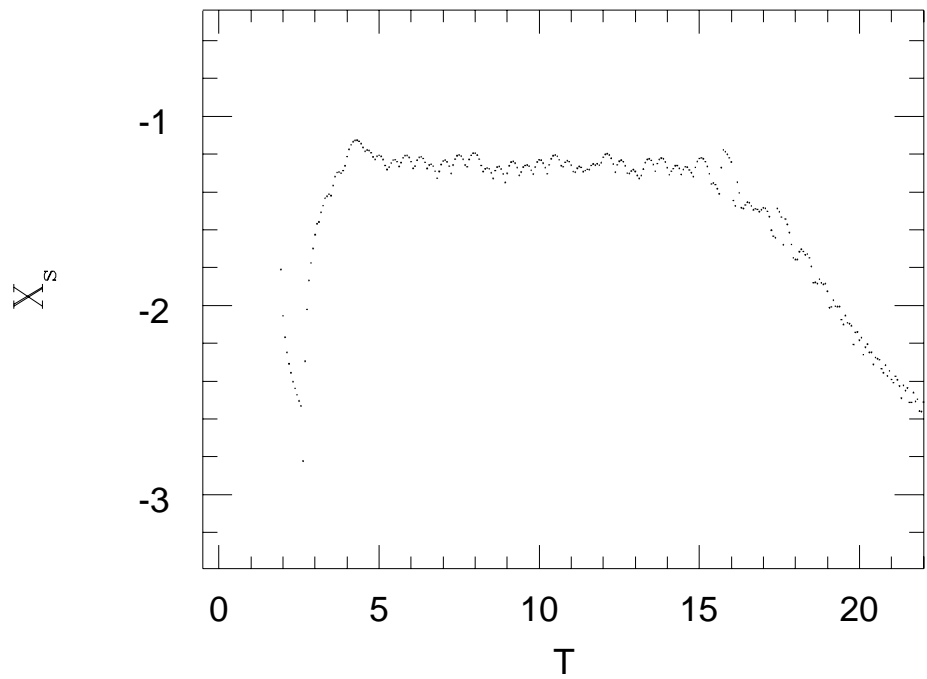
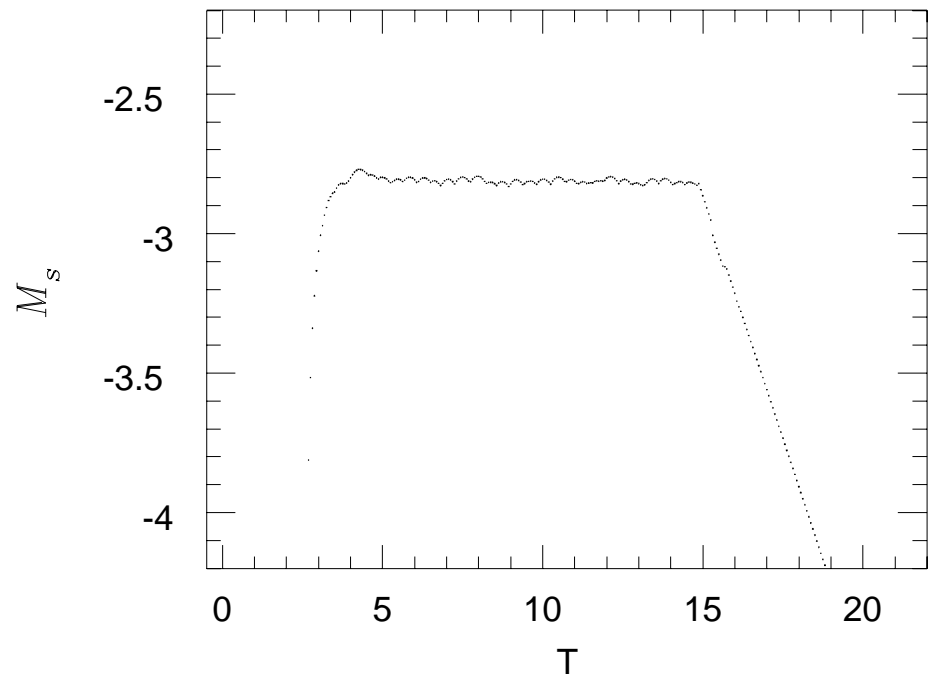
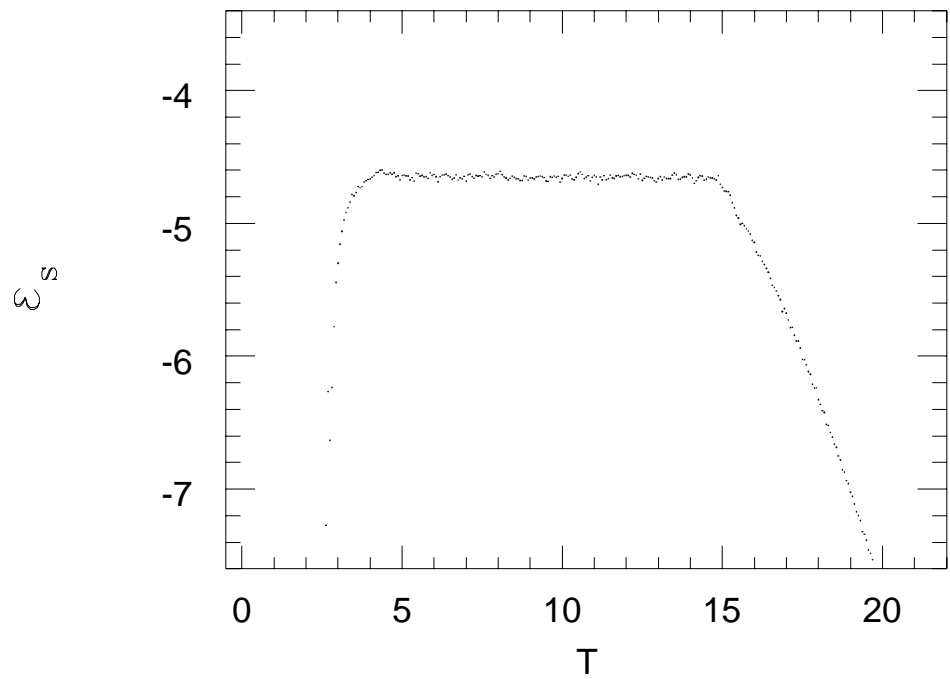
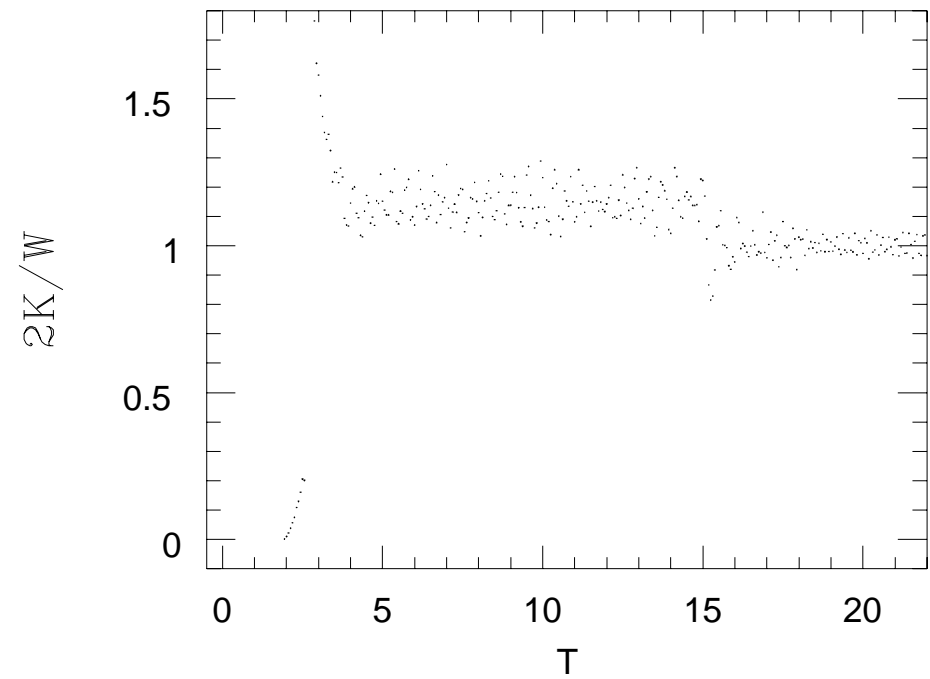
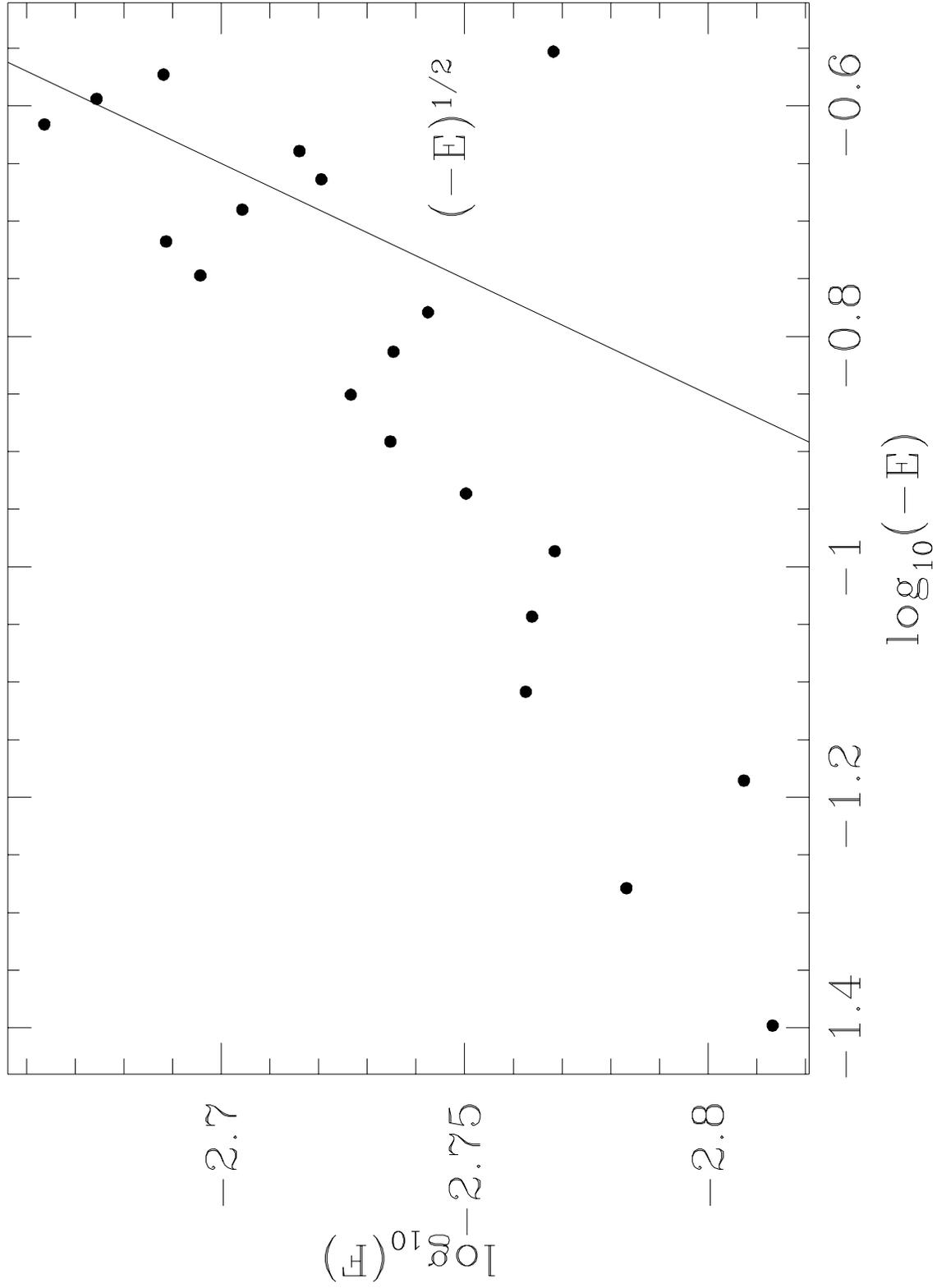
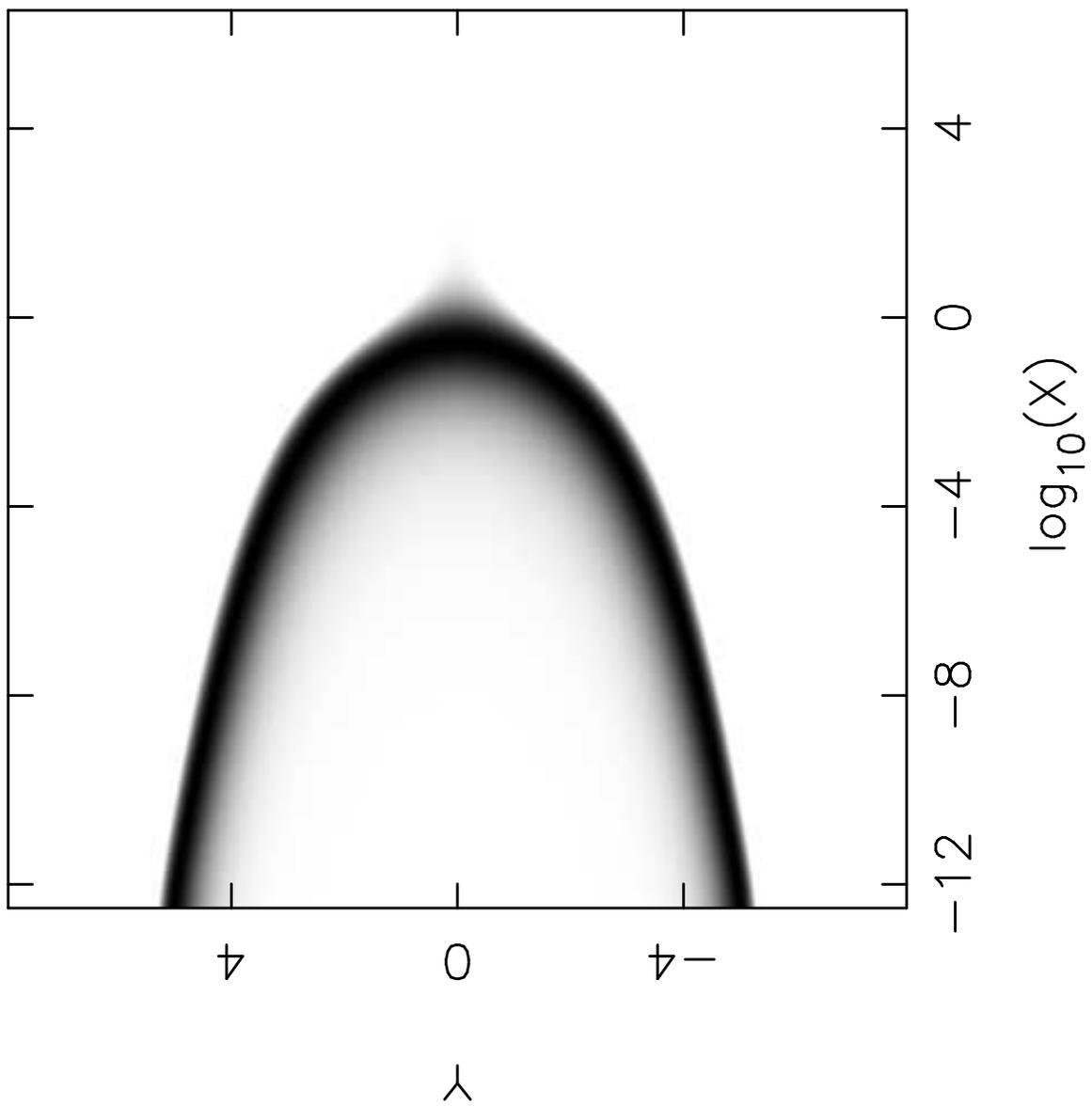
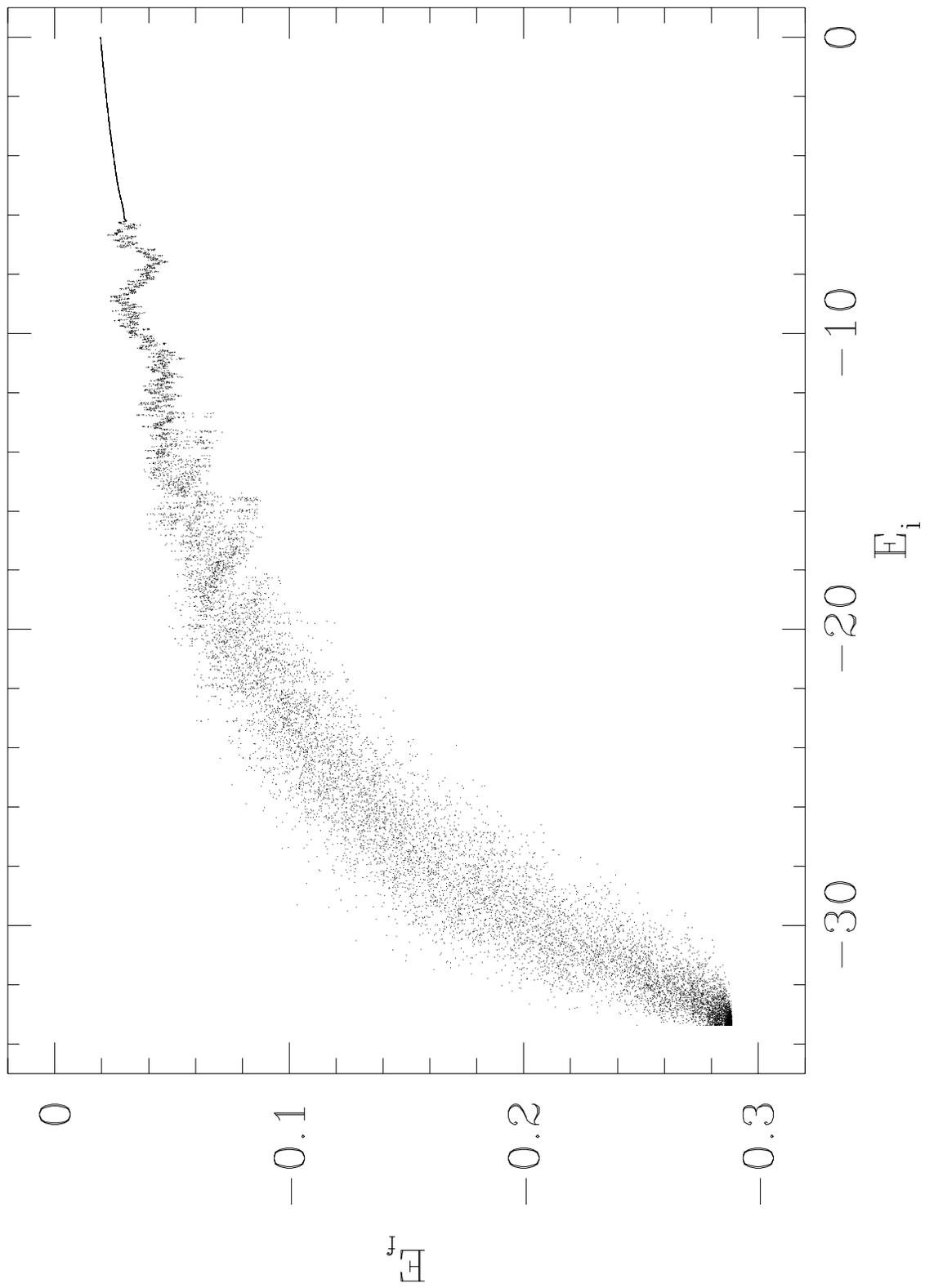
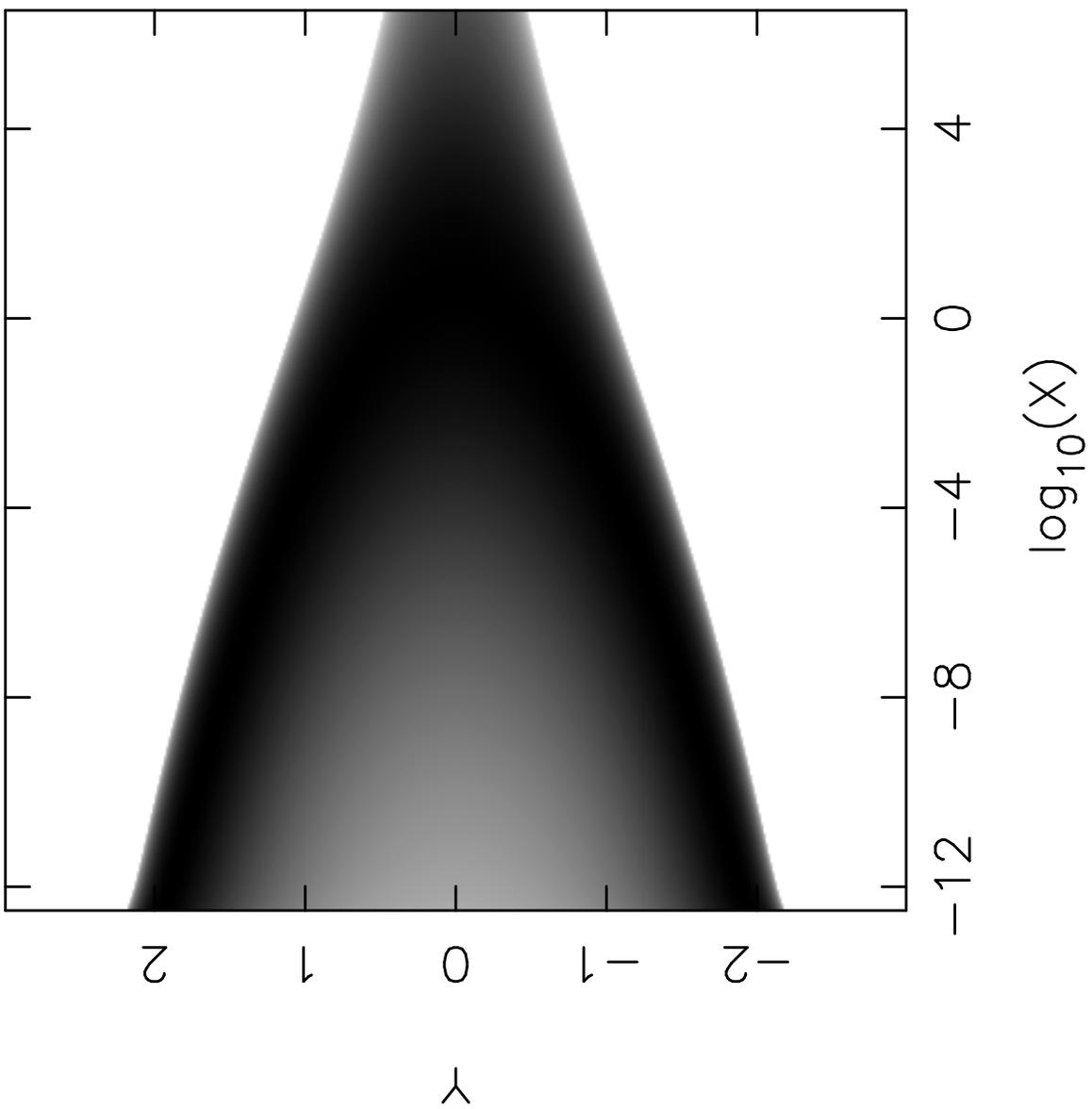
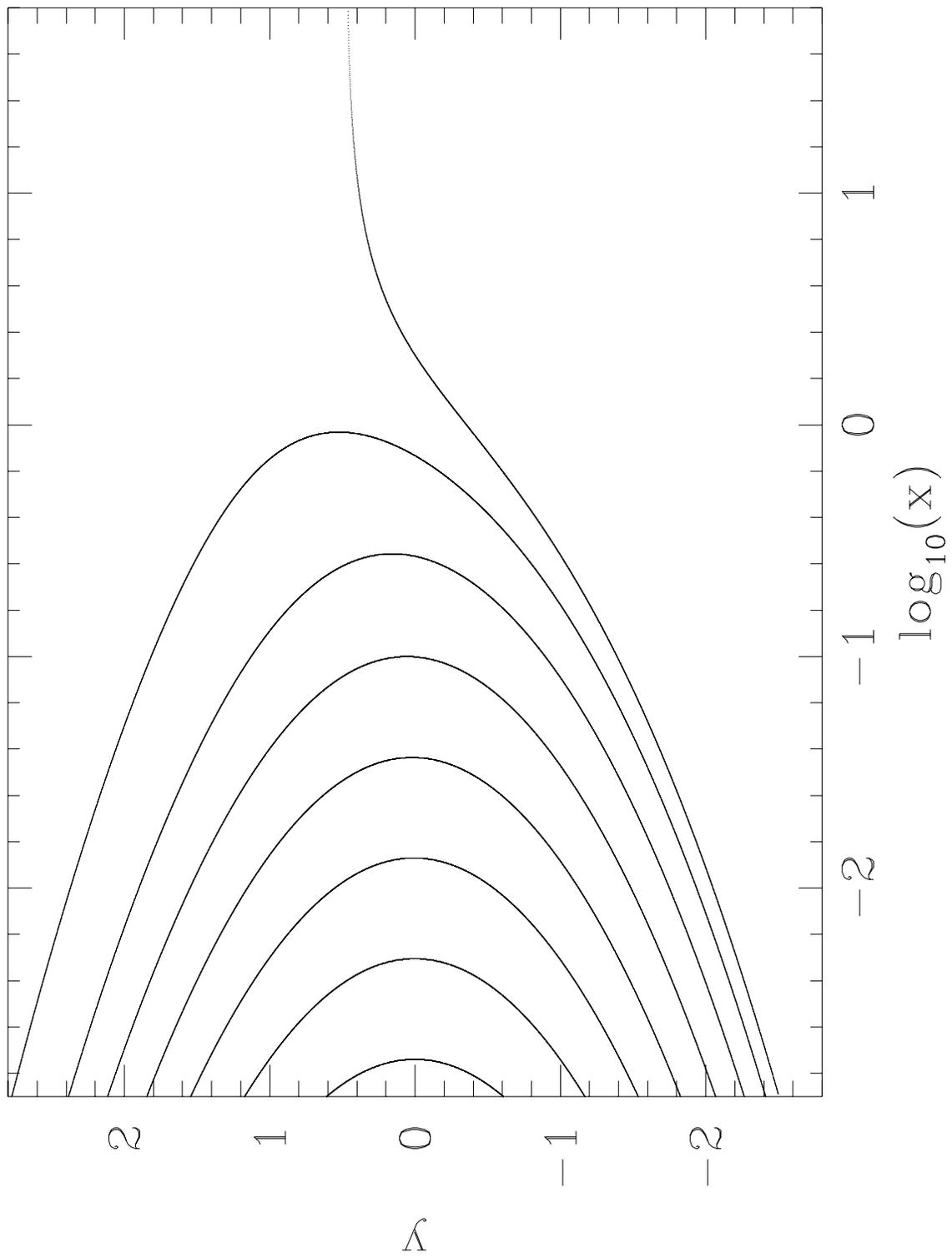
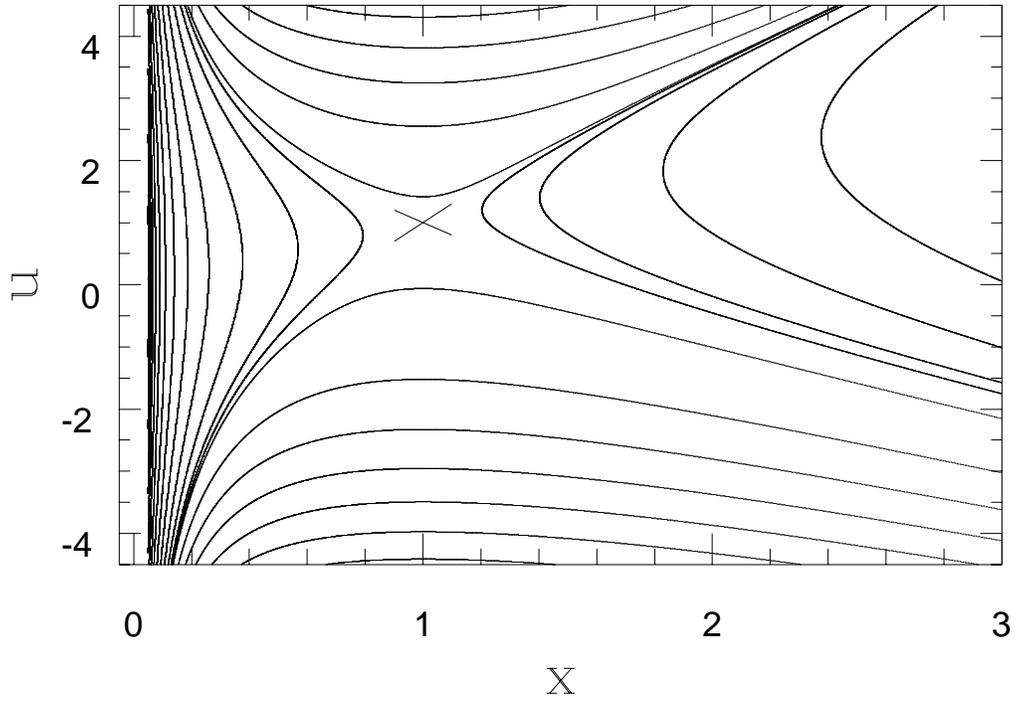
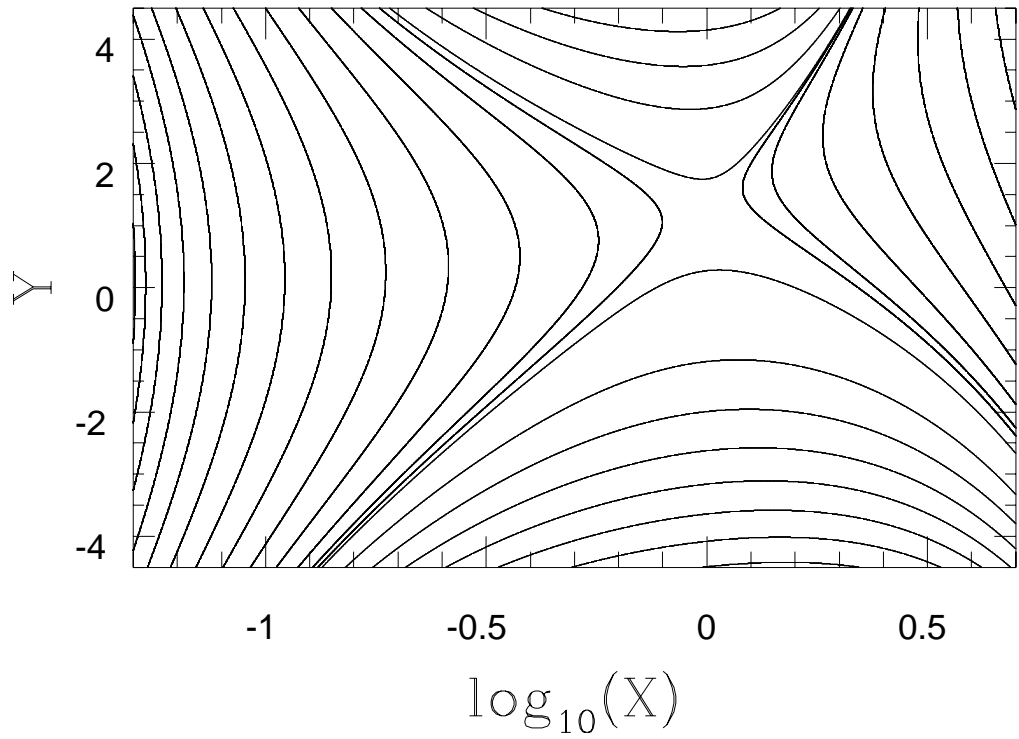
\end{document}